\documentclass[
    aps,prx,twocolumn,
    longbibliography
    reprint,
    superscriptaddress,
    floatfix
]{revtex4-2}

\usepackage[utf8]{inputenc}

\usepackage{amsmath}
\usepackage{amsfonts}
\usepackage{amssymb}
\usepackage{bm}

\usepackage{graphicx}

\usepackage{booktabs}
\usepackage{multirow}

\usepackage[textheight=672pt,textwidth=510pt]{geometry}

\usepackage{xcolor}

\definecolor{linkColor}{RGB}{0,100,150}

\usepackage[colorlinks=true, allcolors=linkColor,pdfborder={0 0 0},pdfencoding = auto]{hyperref}

\DeclareMathOperator{\diag}{diag}
\DeclareMathOperator{\sgn}{sgn}
\let\Im\relax
\DeclareMathOperator{\Im}{Im}
\let\Re\relax
\DeclareMathOperator{\Re}{Re}

\usepackage{bm}
\def\fvec{\bm{f}}
\def\uvec{\mathbf{u}}

\newcommand\kD{k_\mathrm{D}} 
\newcommand\kdD{k_\mathrm{dD}}
\newcommand\kde{k_\mathrm{de}}
\newcommand\kdE{k_\mathrm{dE}}

\newcommand\cDD{c_\mathrm{DD}}
\newcommand\cDT{c_\mathrm{DT}}
\newcommand\cE{c_\mathrm{E}}
\newcommand\md{m_\mathrm{d}}
\newcommand\mde{m_\mathrm{de}}

\newcommand\DD{D_\mathrm{D}}
\newcommand\Dd{D_\mathrm{d}}
\newcommand\Dde{D_\mathrm{de}}

\newcommand\DE{D_\mathrm{E}}

\begin{document}

\raggedbottom

\title{Nonreciprocal pattern formation of conserved fields}

\author{Fridtjof Brauns}
\email{fbrauns@kitp.ucsb.edu}
\affiliation{Kavli Institute for Theoretical Physics, University of California Santa Barbara,
Santa Barbara, California 93106, USA}
\author{M.\ Cristina Marchetti}
\email{cmarchetti@ucsb.edu}
\affiliation{Department of Physics, University of California Santa Barbara, Santa Barbara, California 93106, USA}

\begin{abstract}
In recent years, nonreciprocally coupled systems have received growing attention. Previous work has shown that the interplay of nonreciprocal coupling and Goldstone modes can drive the emergence of temporal order such as traveling waves. We show that these phenomena are generically found in a broad class of pattern-forming systems, including mass-conserving reaction--diffusion systems and viscoelastic active gels. All these systems share a characteristic dispersion relation that acquires a non-zero imaginary part at the edge of the band of unstable modes and exhibit a regime of propagating structures (traveling wave bands or droplets). We show that models for these systems can be mapped to a common ``normal form'' that can be seen as a spatially extended generalization of the FitzHugh--Nagumo model, providing a unifying dynamical-systems perspective. We show that the minimal nonreciprocal Cahn--Hilliard (NRCH) equations exhibit a surprisingly rich set of behaviors, including interrupted coarsening of traveling waves without selection of a preferred wavelength and transversal undulations of wave fronts in two dimensions. We show that the emergence of traveling waves and their speed are precisely predicted from the \emph{local} dispersion relation at interfaces far away from the homogeneous steady state.
Our work thus generalizes previously studied nonreciprocal phase transitions and shows that interfaces are the relevant collective excitations governing the rich dynamical patterns of conserved fields.

\end{abstract}

\maketitle

\section{Introduction}

Multispecies systems with effective cross-interactions that are nonreciprocal have received significant interest in recent years. At the microscale, classical nonreciprocity is intrinsically rooted in the breaking of detailed balance. At a mesoscopic scale, it manifests itself in effective dynamical cross couplings that correspond to non-conservative forces and cannot be obtained as derivatives of a Hamiltonian or free energy. Non-reciprocity is ubiquitous in active and non-equilibrium systems \cite{Fruchart.etal2021}.
It occurs, for instance, in predator-prey systems \cite{Reichenbach.etal2007}, active solids with odd elasticity \cite{Scheibner.etal2020}, protein-based pattern formation \cite{Halatek.etal2018}, mixtures of active and passive particles \cite{You.etal2020}, quorum-sensing active particles \cite{Dinelli.etal2023}, directional interface growth \cite{Pan.DeBruyn1994}, and non-Hermitian quantum systems \cite{Miri.Alu2019}. Such systems can spontaneously organize in dynamical steady states with nontrivial temporal order, such as traveling and oscillating states.

Previous work has examined the effect of nonreciprocity in models of conserved diffusive scalar fields~\cite{You.etal2020,Saha.etal2020,Frohoff-Hulsmann.etal2021}. Nonreciprocal cross-diffusive coupling has been shown to result in the emergence of traveling waves (TWs), whose appearance in systems with conservation laws where fluctuations are expected to decay diffusively is surprising. The directed motion of traveling waves provides a mechanism for the breaking of polar symmetry in system with purely scalar order parameters.  

A similar mechanism is at play in the antagonistic coupling of two groups of flocking agents described for instance by nonreciprocally coupled Toner-Tu equations~\cite{Fruchart.etal2021}. In the absence of coupling, each population  undergoes a phase transition to a state of finite mean motion that spontaneously breaks polar symmetry~\cite{Toner.Tu1995}---the non-equilibrium analog of a finite magnetization in interacting $XY$ spins. When the two populations A and B are coupled anti-reciprocally (A wants to align with B, but B wants to antialign with A) the system is dynamically frustrated and organizes into a state of chase-and-run motion where agents chase each other---a state that breaks chiral symmetry~\cite{Fruchart.etal2021}.
Both sets of results have opened up a flurry of activity on the role of nonreciprocity in dynamical pattern formation~\cite{Bowick.etal2022} and the search for generic models of non-equilibrium transitions from static to time-ordered states~\cite{Frohoff-Hulsmann.Thiele2023}.

Traveling waves are ubiquitous and well understood in nonlinearly dynamical systems with activator/inhibitor couplings. In particular, a prototypical model for oscillations, excitability, and bistability is provided by the FitzHugh--Nagumo (FHN) equations---a set of two coupled nonlinear ODEs originally introduced to describe spike generation in stimulated neurons~\cite{FitzHugh1961,Nagumo.etal1962}.
The spatially extended FHN reaction--diffusion model and its extensions have been studied extensively as prototypical models for the emergence of traveling waves in oscillatory and excitable media \cite{Rocsoreanu.etal2000,Gelens.etal2014}, but a similarly unified description of the origin of time order in systems with conserved quantities is still out of reach. In this paper, we show that a generic minimal model for the transition from static to dynamic patterns in such systems is obtained by coupling the Cahn--Hilliard equation nonreciprocally to a purely diffusive field. The resulting Non-Reciprocal Cahn-Hilliard (NRCH) provides a unified description of a new class of dynamical pattern formation.

In the remainder of this introduction we first present the NRCH model and then summarize the main results of our work.

\subsection{Nonreciprocal Cahn--Hilliard equation}
\begin{figure}[tb]
    \centering
    \includegraphics{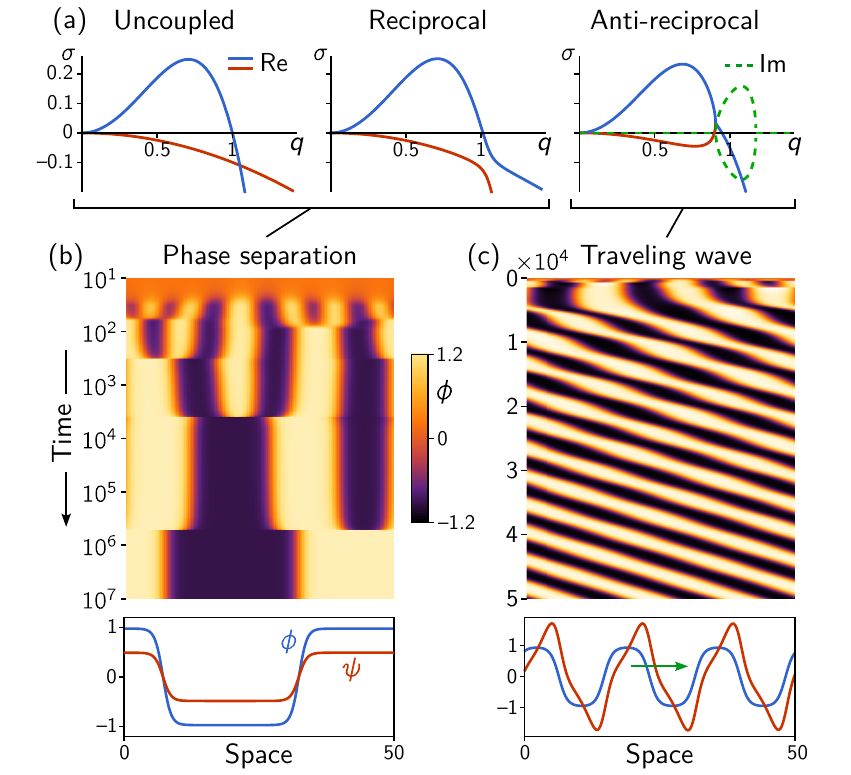}
    \caption{
    (a)~Dispersion relations of the NRCH equations showing the eigenvalue crossing in the uncoupled case ($D_{12} D_{21} = 0$, left) which becomes an avoided crossing for reciprocal coupling (center) and gives rise to a band of propagating modes for anti-reciprocal coupling (right).
    (b), (c)~Kymographs showing the spatiotemporal dynamics in 1D with periodic boundary conditions. (b)~Phase separation and coarsening for reciprocal coupling ($D_{12} = D_{21} = 0.07$).
    (c)~Traveling waves emerge for sufficiently strong anti-reciprocal coupling ($D_{12} = -D_{21} = 0.14$). Notably, coarsening is interrupted for traveling waves.
    [Parameters: $\bar{\phi} = 0, D_{22} = 0.1$; system size $L = 20$. Details on the numerical simulations are provided in Appendix~\ref{app:numerics}. The snapshot profiles in (b) and (c) are from timepoints $t = 10^7$ and $t = 5\times10^4$ respectively.]}
    \label{fig:intro}
\end{figure}

The Cahn--Hilliard equation is a classical model for phase separation of binary fluid mixtures, where two immiscible fluids of conserved mass spontaneously demix. It describes both the equilibrium states and the kinetics of phase separation in terms of a single scalar field that characterizes the conserved volume fraction $\phi$ of one of the two components. 
The minimal nonreciprocal Cahn--Hilliard equations (NRCH) discussed in this work are obtained by coupling $\phi$ to a second conserved and purely diffusive field $\psi$. The coupled dynamics of the two fields is given by 
\begin{subequations} \label{eq:NRCH}
\begin{align}
    \partial_t \phi(x,t) &= \nabla^2 (D_{11} \phi + D_{12} \psi + \phi^3 - \kappa \nabla^2 \phi) \,, \label{eq:NRCH-phi}\\
    \partial_t \psi(x,t) &= \nabla^2 (D_{21} \phi + D_{22} \psi)\,, \label{eq:NRCH-psi}
\end{align}
\end{subequations}
with diffusion coefficients $D_{ij}$. We demand $D_{22} > 0$ to ensure stability of the $\psi$ field. The $\kappa$-term (with $\kappa > 0$) stabilizes the phase-separating $\phi$-field at short scales and thus controls the interface width and tension.
Phase separation can occur through a spinodal instability when $D_{11}<0$. 
This is evident by examining the linear dynamics of fluctuations of the conserved fields $\phi$ and $\psi$ from their homogeneous values. Since these fields are conserved, the dynamics of fluctuations is controlled by soft or hydrodynamic modes, defined as those where a fluctuation of wavenumber $q$ decays (or grows) at a rate $\sigma(q)$, with $\lim_{q\to 0} \Re[\sigma(q)]=0$. A familiar example is  diffusion, where a density fluctuations can only decay by redistributing material throughout the system. 
Indeed, in the absence of cross couplings, fluctuations in $\psi$ decay diffusively, while fluctuations in $\phi$ exhibit the characteristic dispersion relation of a spinodal instability [see Fig.~\ref{fig:intro}(a)], with largest growth at a characteristic length scale controlled by $|D_{11}| $ and $\kappa$.

The minimal nonlinearity ($\phi^3$) saturates the pattern amplitude, resulting in stable bulk phase separated regions at long times [Fig.~\ref{fig:intro}(b)].

Cross-diffusion couples the two hydrodynamic modes, which becomes apparent through the interaction of the eigenvalue branches in the linear fluctuation spectrum (dispersion relation).
For cross-diffusivities with equal signs, the branches avoid crossing and remain real [Fig.~\ref{fig:intro}(a)].
By contrast, nonreciprocal cross-diffusivities ($D_{12}\not= D_{21}$) cause the dispersion relation branches to cross at a point where the corresponding eigenvectors align, resulting in the appearance of an imaginary part [green dashed lines in Fig.~\ref{fig:intro}(a)].
This \emph{mode coalescence} due to nonreciprocal couplings transforms the static phase separated state into traveling [Fig.~\ref{fig:intro}(c)] or oscillating domains~\cite{You.etal2020,Saha.etal2020} through a generic mechanism discussed in detail below. While this model and more complex ones consisting of two coupled Cahn--Hillard equations have been studied before~\cite{You.etal2020,Saha.etal2020,Frohoff-Hulsmann.etal2021}, previous work has some important limitations. Specifically, Ref.~\cite{You.etal2020} has examined Eqs.~\eqref{eq:NRCH} only for small systems near the onset of instability. Refs.~\cite{Saha.etal2020,Frohoff-Hulsmann.etal2021} have considered more complex equations, which renders an exhaustive analysis and intuitive understanding of the dynamics difficult.

\subsection{Summary of results and outline}

Our work significantly extends previous findings in a number of ways. It shows that, despite their apparent simplicity, Eqs.~\eqref{eq:NRCH} provide deep insight into a broad class of pattern forming systems and
exhibit several remarkable behaviors that had not been previously reported, including interrupted coarsening of traveling waves without wavelength selection and undulating interfaces. 

\paragraph{A generic model of traveling patterns in extended systems.} First, we  show that Eqs.~\eqref{eq:NRCH} provide a unifying minimal model for a non-equilibrium phase transition from static to time-ordered states in extended systems with conservation laws.
Specifically, \emph{antagonistic} cross-diffusion causes spatial inhomogeneities in the two fields to undergo a chase-and-run dynamics, until they settle in a state with a common velocity. This is signaled by the mode coalescence via the merging of the real part of the two hydrodynamic modes' growth rates and the simultaneous emergence of a finite imaginary part indicating propagating waves [see Fig.~\ref{fig:LSA}(b-e)]. This structure of the dispersion relation is a distinct signature of temporal organization, hence provides a criterion for identifying this new class of dynamical pattern formation. Furthermore, the hydrodynamic nature of the fluctuations guarantees that for strong enough nonreciprocity the merging will occur for all parameter values, hence it is generic.
Connecting with the language of dynamical systems, the transition of the decoupled field $\phi$ from a homogeneous well-mixed state to a stationary phase  composed of dilute and dense regions occurs via a pitchfork bifurcation~\cite{StrogatzBook} of the pattern amplitude. Nonreciprocal cross couplings drive a second transition to a time-dependent state that breaks parity and time reversal symmetry, where the demixed domains (``droplets'') travel at constant velocity. The domains organize into periodic wave trains that we refer to as traveling waves. This transition is related to ones previously observed in non-conserving reaction--diffusion systems \cite{Malomed.Tribelsky1984,Coullet.etal1989a,Fauve.etal1991} and is known as a drift-pitchfork bifurcation~\cite{Kness.etal1992}.
The same bifurcation underlies the transition to collective actuation in an active solid \cite{Baconnier.etal2022}.

\paragraph{A criterion for identifying this new class of dynamical pattern formation.} We identify the spatially localized coalescence of hydrodynamic modes arising from the conservation laws shown in Fig.~\ref{fig:LSA}(b-e) as a generic mechanism for temporal organization in these systems. The features of the linear dispersion relation highlighted in Fig.~\ref{fig:LSA}(b) thus provide a criterion for identifying this new class of dynamical pattern formation. 
The intersection of the two branches of the dispersion relation is associated with a non-diagonalizable form of the matrix that governs the linear dynamics of fluctuations and degenerate eigenvectors. This mechanism is analogous to the one responsible for the onset of chiral states in antagonistic flocking models and is referred to as an exceptional point in the corresponding phase diagram~\cite{Fruchart.etal2021}. 
There is, however, an important difference between the two systems.  In flocking models the velocity order parameter is not conserved. Instead, the spontaneously broken polar symmetry of the flocking state is associated with the emergence of a Goldstone mode with relaxation rate that vanishes at long wavelength.
Chiral states that spontaneously select handedness occur when the Goldstone mode coalesces with the relaxational mode describing fluctuations of the homogeneous steady state.~\cite{Fruchart.etal2021}. The coalescence therefore happens \emph{globally}.
In contrast, for the mass-conserving systems studied here, mode coalescence is spatially \emph{localized} at the interfaces of the phase separated patterns and the emergence of traveling states that break polar symmetry occurs generically for all parameter values.
The interfaces effectively ``self-tune'' to the neutrally stable mode at the right edge of the band of unstable modes. This behavior is manifest in the characteristic form of the dispersion relation [Fig.~\ref{fig:LSA}(b--e)] that identifies the class of systems captured by the minimal model, Eq.~\eqref{eq:NRCH}.
Importantly, this allows us to predict the speed of the traveling waves by a \emph{local} dispersion relation at the interface, even far from the homogeneous steady state. This points towards a vantage point for tackling highly nonlinear systems by linearizing locally and using the presence of conservation laws as previously proposed in Refs.~\cite{Halatek.Frey2018,Brauns.etal2020}.

\paragraph{A unifying model of traveling patterns.}
We explicitly demonstrate that the mode-coalescence route to spatiotemporal order arises naturally in a broad class of pattern-forming systems that can all be recast in the NRCH framework. We show this explicitly for one-dimensional realizations of  non-equilibrium systems previously considered in the literature, including mixtures of active and passive particles \cite{You.etal2020}, mass-conserving reaction--diffusion systems \cite{Brauns.etal2021b,John.Bar2005a,Frohoff-Hulsmann.Thiele2023}, and active gel models \cite{Radszuweit.etal2013,Weber.etal2018}.
In all these systems an analysis of the linear dispersion relation at the inflection point of the interface between two phase separated regions using a method introduced in Ref.~\cite{Brauns.etal2020}, provides an expression predicting the interface speed---a result that generally requires an analysis deep in the nonlinear region.

\paragraph{A new mechanism for interrupted coarsening.}
We show that, although the equations contain only a single explicit length scale, corresponding to the width of the interface, {the traveling droplets exhibit interrupted/arrested coarsening. It is known that the introduction of a second length scale, as arising for instance from broken mass conservation, can arrest coarsening~\cite{Glotzer.etal1995,Cates.etal2010,Curatolo.etal2020,Brauns.etal2021}) Other mechanisms that have been shown to interrupt coarsening resulting in micro-phase-separated states include elasticity~\cite{Rosowski.etal2020,Qiang.etal2023}, the coupling to chemical reactions~\cite{Zwicker2022}, and long-range interactions~\cite{Liu.Goldenfeld1989,Ohta.Kawasaki1986}. The mechanism that stabilizes finite wavelength patterns in the NRCH model is, however, more subtle. In the initial stages of the phase separation droplets travel at different speed and the system continues to coarsen via collisions of traveling droplets in a chase-and-run mode until all droplets have achieved the same stable velocity and coarsening stops. As a result, one finds that the system can select stable  traveling droplets with  a broad range of wavelengths---ranging from the interface width to the system size---depending on the initial conditions.}

\paragraph{New dynamical patterns in two-dimensional systems.}
In two-dimensional (2D) systems, interfaces can become unstable leading to undulations of the wave fronts. This interfacial instability gives rise to a rich variety of patterns including spatiotemporal chaos. Notably, since undulations propagate along the interfaces, i.e.\ transversally to the wave fronts, they break chiral symmetry in addition to polar symmetry.

\paragraph{Role of boundary conditions.}
No-flux boundaries break translation invariance and can arrest traveling waves in one dimension (1D). For sufficiently strong anti-reciprocal coupling, stationary solutions cease to exist giving rise to standing waves in 1D and to periodically sloshing structures in 2D. In 1D the onset of the transition to the standing wave regime can be read off from a simple geometric condition in the phase portrait and the full phase diagram can be predicted by linear stability analysis and simple graphical constructions. 

\paragraph{Outline.}
The remainder of this paper is organized as follows. In Section~\ref{sec:FHN} we revisit the FitzHugh--Nagumo model to highlight the role of nonreciprocity in driving temporal oscillations (limit cycles). In Section~\ref{sec:NRCH} we introduce the minimal NRCH model, carry out a linear stability analysis, and show that a regional stability analysis in the interfacial region provides an excellent estimate for the speed of traveling waves. After discussing the arrest of coarsening and wavelength selection in 1D in Section~\ref{sec:coarse}, we extend the numerical work to 2D, where a new state of undulating traveling waves is observed (Section ~\ref{sec:2D}). In Section~\ref{sec:systems} we show that a number of recently studied physical systems can be mapped onto our minimal NRCH model. Details of the mapping are given in Appendix~\ref{app:MCRD} for reaction-diffusion systems, with specific application to the dynamics of the Min-protein system of \emph{E.\ coli}, and in Appendix~\ref{app:poroelastic} for active gel models of active poroelastic media. We conclude the body of the paper with a brief discussion and outlook in Section~\ref{sec:discussion}.

\section{Revisiting the FitzHugh--Nagumo model}
\label{sec:FHN}

\begin{figure}
    \centering
    \includegraphics{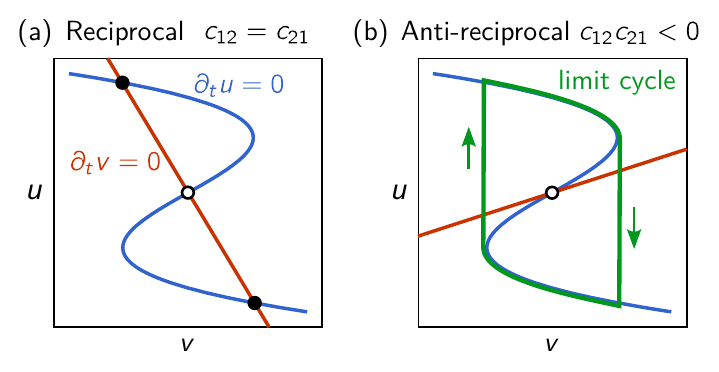}
    \caption{Phase portrait of the FHN model~\eqref{eq:FHN} with $a = 0$. (a) For reciprocal coupling ($c_{12} = c_{21}$), the $v$-nullcline is always sloped such that the system is bistable. (b) For sufficiently strong anti-reciprocal coupling, limit-cycle oscillations emerge.}
    \label{fig:FHN}
\end{figure}

We start with a minimal example that places nonreciprocity in a dynamical systems framework. To this end, we briefly revisit the classic FitzHugh--Nagumo (FHN) model \cite{FitzHugh1961,Nagumo.etal1962}
\begin{subequations} \label{eq:FHN}
\begin{align} 
    \partial_t u(t) &= u - u^3 + c_{12} v\,, \label{eq:FHN-u}\\
    \partial_t v(t) &= a - b v + c_{21} u\,, \label{eq:FHN-v}
\end{align}
\end{subequations}
which serves as a prototypical minimal model for excitability, bistability and oscillations in a broad variety in physical and biological systems. In the biological context, $u$ is called an activator and $v$ an inhibitor when $c_{12} < 0$ and $c_{21} > 0$.
When $c_{12} = c_{21} = c$, Eq.~\eqref{eq:FHN} can be derived as relaxational dynamics in the free energy landscape $f(u, v) = -\frac12 u^2 + \frac14 u^4 - c u v - av + \frac{b}{2} v^2$. 

{It is useful to analyze the behavior in the language of dynamical systems \cite{StrogatzBook,IzhikevichBook} by plotting the nullclines in the $(u,v)$ phase space, defined as the solutions of $\partial_tu=0$ and $\partial_tv=0$ (Fig.~\ref{fig:FHN}(a,b)). The intersection of the nullclines determines the fixed points of the system.} In the reciprocal case {the system is bistable: there is an unstable fixed point at $u=v=0$ and two stable fixed points [black disks in Fig.~\ref{fig:FHN}(a)].} The dynamics cannot exhibit limit cycle oscillations, as is manifest from the phase portrait Fig.~\ref{fig:FHN}(a).
By contrast, oscillations appear when the coupling is sufficiently nonreciprocal $c_{12} \neq c_{21}$. In the regime of separated timescales, $|a|, b, |c_{12}| \ll 1$, the limit cycle oscillations can be constructed geometrically in the phase portrait as relaxation oscillations which periodically switch the state of the system between the two stable steady state branches of the fast $u$-dynamics [see Fig.~\ref{fig:FHN}(b)].
Thus, the FHN model exemplifies that a dynamical systems perspective allows one to understand geometrically how oscillations emerge from nonreciprocal coupling.
Such insight is useful because, unlike the amplitude equation formalism commonly employed to study pattern formation \cite{Cross.Hohenberg1993}, it is not restricted to the vicinity of the onset of oscillations. 
In the following, we shall seek this kind of insight for the dynamics of Eqs.~\eqref{eq:NRCH}.

\section{A prototypical model for nonreciprocally coupled conserved fields}
\label{sec:NRCH}

To go from the ``well-mixed'' setting described by a few degrees of freedom to a spatially extended system, it is useful to first consider just the bistable dynamics of $u$, i.e.\ Eq.~\eqref{eq:FHN-u} with $c_{12} = 0$. Bistable dynamics can be cast as relaxational dynamics in a double-well potential.
The minimal prescription for a spatially extended system is to supplement the double-well free energy density by the lowest order term of a gradient expansion, $\frac{\kappa}{2} (\nabla u)^2$, penalizing interfaces. The relaxational dynamics can be either non-conserved in which case one obtains the Allen--Cahn equation \cite{Allen.Cahn1975} or conserved giving the Cahn--Hilliard equation \cite{Cahn.Hilliard1958}, i.e.\ Eq.~\eqref{eq:NRCH-phi} with $D_{12} = 0$. (These equations are also referred to as Model~A and Model~B, respectively \cite{Hohenberg.Halperin1977}.) We can now apply the same logic to the FHN equations, by deriving the relaxational dynamics from the free energy density $f + \frac{\kappa}{2} (\nabla u)^2$ (yielding reciprocal dynamics) and then allowing the coupling coefficients to become nonreciprocal. 
In the non-conserved case, one arrives at the well-known FHN reaction--diffusion equations which are a prototypical model for oscillatory and excitable media, exhibiting phase and trigger waves \cite{Mikhailov1990}. 
In the conserved case, one obtains the NRCH model, Eqs.~\eqref{eq:NRCH}.
Table~\ref{tab:FHN-CH} provides an overview of the relationships between different models discussed above and places the NRCH model in a systematic classification scheme.
To emphasize that the variables are fields and avoid confusion with the FHN ODEs~\eqref{eq:FHN}, we have changed notation to the fields $\phi$, $\psi$ and the diffusion coefficients $D_{ij}$.

Before analyzing the emergent behavior arising from Eqs.~\eqref{eq:NRCH} in further detail, let us get a few preliminaries out of the way. 
The spatial averages of $\bar{\phi} = \langle \phi \rangle$ and $\bar{\psi} = \langle \psi \rangle$, are conserved and are therefore control parameters of the system.
However, because the Eqs.~\eqref{eq:NRCH} are invariant under addition of a global constant to $\psi$, we can set $\bar{\psi} = 0$ without loss of generality.
Moreover, we have scaled the coefficient in front of the $\phi^3$ term to unity which is always possible by rescaling the field $\phi$.
We further choose length and timescales such that $\kappa = 1$, $D_{11} = -1$.
Observe that one can rescale $\psi$ such that either $D_{21} = D_{12}$ (reciprocal case) or $D_{21} = -D_{12}$ (anti-reciprocal case).
In other words, only the product $D_{12} D_{21}$ is relevant for the dynamics. 
We therefore introduce the signed geometric mean of the cross-diffusivities,
\begin{equation}
    \tilde{D} := \sgn(D_{12} D_{21}) \sqrt{|D_{12} D_{21}|} \,,
\end{equation}
as the parameter controlling the strength and reciprocity of cross-diffusive coupling throughout the paper. That is, we set $D_{12} \to \tilde{D}$, $D_{21} \to |\tilde{D}|$ in Eq.~\eqref{eq:NRCH}.

Finally, we note that by discretizing Eqs.~\eqref{eq:NRCH} in the elementary setting of two well-mixed, diffusively coupled compartments, one obtains equations for the mass differences between the two compartments whose form recovers the FHN equations (see Appendix~\ref{app:mapping-to-FHN}). Static and standing wave patterns correspond to bistability and limit-cycle oscillations in the FHN model, respectively.

\begin{table}[t]
    \centering
    \begin{tabular*}{\linewidth}{@{\extracolsep{\fill}}llll@{}}
        \toprule
        \multirow{2}{*}{Local} & & \multirow{2}{*}{Bistability} & Oscillations, \\
        &&& Excitability\\
        \midrule
        \multirow{2}{*}{Spatial $\;\bigg\{$} & Non-conserved & Allen--Cahn & FHN-RD \vspace{0.3em}\\
        & Conserved & Cahn--Hilliard & NRCH: Eq.~\eqref{eq:NRCH} \\
        \bottomrule
    \end{tabular*}
    \caption{``Local'' dynamics and their generalizations to spatially extended systems. FHN-RD is the classic FHN reaction--diffusion model for oscillatory and excitable media obtained by supplementing Eq.~\eqref{eq:FHN} by diffusion of $u$ and $v$.}
    \label{tab:FHN-CH}
\end{table}

\subsection{Linear stability analysis}

We begin with a linear stability analysis of the homogeneous steady states. Notably, this analysis will later prove useful beyond the usual setting of linearization around a \emph{global} steady state. Instead, we will use a \emph{local} dispersion relation---calculated at the interfaces of highly nonlinear patterns---to predict the onset and speed of traveling waves.

\begin{figure}[t]
    \includegraphics{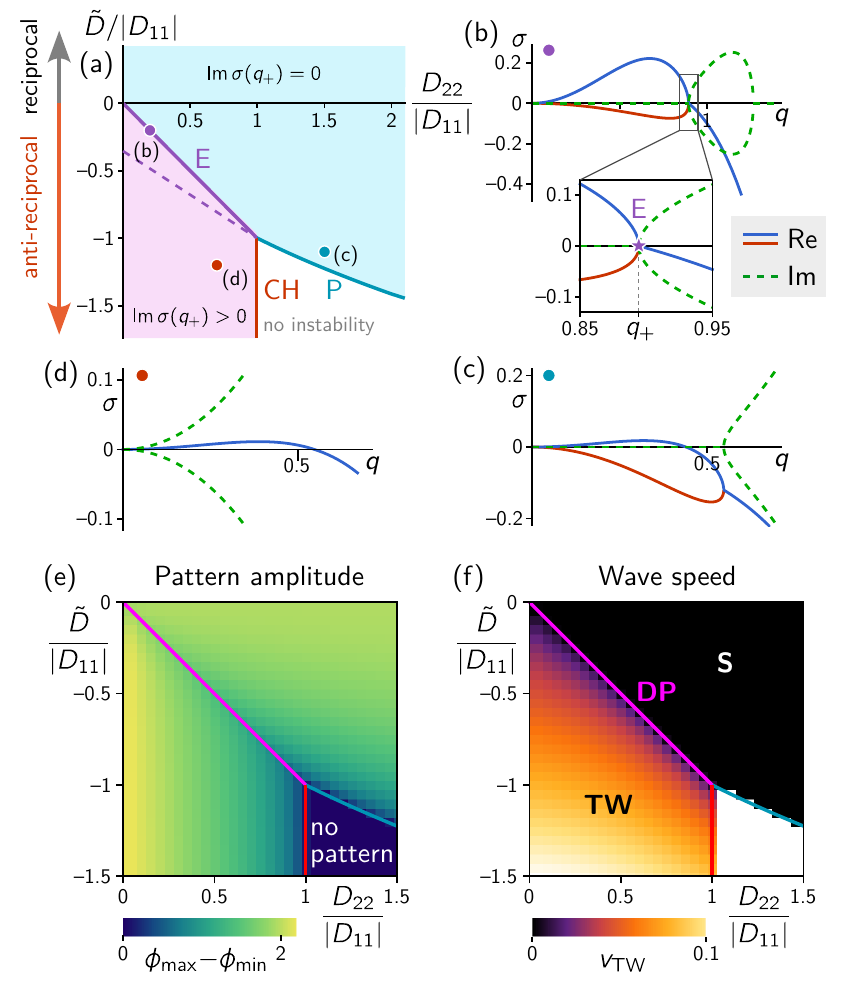}
    \caption{
    (a) Linear stability diagram in the $D_{22}$-$\tilde{D}$ parameter plane for an equal mixture ($\bar{\phi} = 0$); for unequal mixtures $\bar{\phi} \neq 0$, see App.~\ref{app:unequal-mix}. 
    Along the solid purple line (E) an ``exceptional point'' appears in the dispersion relation, where the band of unstable modes touches the band of propagating modes [see (b)]. CH and P indicate the ``conserved Hopf'' and pitchfork bifurcations bounding the regime where pattern formation is suppressed by fast $\psi$ diffusion ($D_{22} > -D_{11}$).
    Below the dashed purple line, the fastest growing mode in the dispersion relation is propagating. 
    (b--d)~Dispersion relations in different regimes: (b)~exceptional point (see inset); (c)~near the pitchfork bifurcation; (d)~near the conserved Hopf bifurcation, where all unstable modes propagate [$\Im \sigma(q \to 0) \sim q^2$].
    (e), (f)~Pattern amplitude and wave speed as a function of $D_{22}$ and $\tilde{D}$. Stationary patterns (S) transition to traveling waves (TW) in a drift-pitchfork (DP) bifurcation at $\tilde{D} = -D_{22}$ which lies exactly along the line of exceptional points [see (a)]. The onset of patterns at the conserved Hopf (CH) and pitchfork (P) bifurcations is supercritical for the case equal mixture case $\bar{\phi} = 0$ shown here.
    [System size $L = 100$ in (e) and (f).]
    }
    \label{fig:LSA}
\end{figure}

Linearizing Eqs.~\eqref{eq:NRCH} around a homogeneous steady state $(\phi, \psi) = (\phi_0, \psi_0)$ \footnote{For the common linear stability analysis of the global homogeneous steady state of the system, $(\phi_0, \psi_0) = (\bar{\phi}, \bar{\psi})$. We will generalize this analysis to a setting where $(\phi_0, \psi_0)$ are local densities.} for perturbations of the form $e^{iqx + \sigma t}$ yields the Jacobian
\begin{equation} \label{eq:Jacobian}
    J = -q^2 \begin{pmatrix}
        D_{11} + 3 \phi_0^2 + \kappa q^2 & \tilde{D} \\
        |\tilde{D}| & D_{22}
    \end{pmatrix} .
\end{equation}
In the uncoupled case ($\tilde{D} = 0$), the dispersion relation has two independent branches given by the diagonal entries in the Jacobian {[see Fig.~\ref{fig:intro}(a), left]}. A band of unstable modes $[0, q_+]$ emerges in the first branch when $D_{11} < -3\phi_0^2$, where $q_+^2 = (-D_{11} - 3\phi_0^2)/\kappa$. This is the well-known spinodal decomposition instability that drives phase separation.
Cross-diffusive couplings cause the branches of the dispersion relation to interact near their intersection point, giving rise to either an avoided crossing in the reciprocal case $\tilde{D} > 0$ {[see Fig.~\ref{fig:intro}(a), center]} or a band of propagating modes ($\Im \sigma \neq 0$) in the anti-reciprocal case $\tilde{D} < 0$; see {Fig.~\ref{fig:intro}(a), right and } Fig.~\ref{fig:LSA}(c--e).
For sufficiently strong anti-reciprocal coupling, the band of propagating modes begins to overlap with the band of unstable modes $[0, q_+]$.
Some simple algebra (and introducing the shorthand $d_{11} = D_{11} + 3\phi_0^2$) yields
\begin{subequations} 
\begin{align}
    q_+ &= \sqrt{- (d_{11} + D_{22})/\kappa}\,, \label{eq:q+TW}\\
    \Im \sigma(q_+) &= \pm q_+^2 \sqrt{- \tilde{D} |\tilde{D}| - D_{22}^2}\,, \label{eq:Im}
\end{align}
if $-\tilde{D} > D_{22}$ and 
\begin{align} 
    q_+ &= \textstyle \sqrt{- (d_{11} - \frac{\tilde{D} |\tilde{D}|}{D_{22}}) / \kappa}\,, \label{eq:q+stat}\\
    \Im \sigma(q_+) &= 0\,,
\end{align}
\end{subequations}
if $-\tilde{D} \leq D_{22}$.
Unstable modes exist only when the expressions under the square root in $q_+$ is positive, which sets the boundaries of the linearly unstable parameter regime, demarcated by ``conserved-Hopf'' (CH) and pitchfork (P) bifurcations in Fig.~\ref{fig:LSA}(a).
Outside this regime, fast $\psi$ diffusion ($D_{22} > -D_{11}$) suppresses pattern formation. The term ``conserved-Hopf'' instability was coined in Ref.~\cite{Frohoff-Hulsmann.Thiele2023} to describe a long-wavelength instability with propagating modes, giving rise to traveling waves with a finite speed at onset \cite{You.etal2020,Frohoff-Hulsmann.Thiele2023} (analogous to the onset of oscillations with finite frequency in a Hopf bifurcation \cite{StrogatzBook}).
At the pitchfork bifurcation, stationary patterns emerge with an amplitude that scales as $\Delta^{1/2}$ where $\Delta$ is the parameter distance from the bifurcation [see Fig.~\ref{fig:stat-state-construction}(d)].
Both these bifurcations are supercritical for $\bar{\phi} = 0$, as can be seen from the pattern amplitude going to zero at the bifurcation [Fig.~\ref{fig:LSA}(e)]. For $\bar{\phi} \neq 0$ they become subcritical (see Appendix~\ref{app:unequal-mix}).

For $\tilde{D} = -D_{22}$, the marginal mode $q_+$ touches the band of propagating modes.
At this point, the Jacobian has two vanishing eigenvalues and is non-diagonalizable, i.e.\ its eigenvectors coincide. This marks an exceptional point \cite{Fruchart.etal2021,Miri.Alu2019}.
Note that an exceptional point has codimension two, meaning that two parameters need to be tuned for it to occur. Here, one of these parameters is the wavenumber $q$. This implies that only one \emph{control parameter} of the system, e.g.\ a cross-diffusion coefficient, needs to be tuned for an exceptional point to appear in the dispersion relation.
(Additional linear stability diagrams in terms of $D_\pm = (D_{21} \pm D_{12})/2$ and in the $D_{11}$-$D_{22}$ plane are shown in Fig.~\ref{fig:Dm-Dp-diagrams} in Appendix~\ref{app:Dplus-Dminus}.)

A puzzling observation in previous literature has been that traveling waves emerge when $\Im \sigma(q_+) \neq 0$, even if the fastest growing mode of the dispersion relation, corresponding to wavenumber $q_\mathrm{max}$ is not propagating (i.e.\ has vanishing imaginary part) \cite{Perelson.etal1986,Zhao.etal2023,Weber.etal2018,Radszuweit.etal2013}.
Indeed numerical simulations show that $\Im \sigma(q_+)$ provides a precise criterion for the onset of traveling waves in a drift-pitchfork (DP) bifurcation \cite{Kness.etal1992} [Fig.~\ref{fig:LSA}(f)].
In such a bifurcation, a stationary pattern becomes unstable and starts propagating (or ``drifting'') and the propagation speed scales as 
$\Delta^{1/2}$ with the distance $\Delta$ from the bifurcation. In other words, the propagation speed undergoes a pitchfork bifurcation---hence the name.
This is an important difference to the CH instability, where traveling waves emerge with a finite speed.
The emergence of traveling waves in DP and CH bifurcations spontaneously breaks parity symmetry in 1D (polary symmetry in higher spatial dimensions).

The prediction of the DP bifurcation from the exceptional point in the dispersion relation is remarkable from the standpoint of weakly nonlinear analysis where one expects traveling waves only when the fastest growing mode is propagating (i.e.\ when $\Im \sigma(q_\mathrm{max}) \neq 0$). 
To resolve this puzzle, we first analyze the fully nonlinear stationary patterns and then use the \emph{local} dispersion relation at its interfaces to understand the transition to traveling waves.

\subsection{Generalized Maxwell construction for stationary patterns}
 \label{sec:stat-state}

Weakly nonlinear analysis, as performed in Ref.~\cite{You.etal2020}, is restricted to small systems of size $L \simeq 2 \pi / q_+$ where only the system-size mode is unstable.
To overcome this limitation, i.e.\ investigate the fully nonlinear traveling waves in a large system, we first take a closer look at the stationary patterns by generalizing the Maxwell construction for phase separation. Stationary patterns will form the basis for a quantitative understanding of the onset and speed of traveling waves.
We focus on one spatial dimension in this and the subsequent section. Two-dimensional patterns will be briefly investigated in Sec.~\ref{sec:2D}.

\begin{figure}
    \centering
    \includegraphics{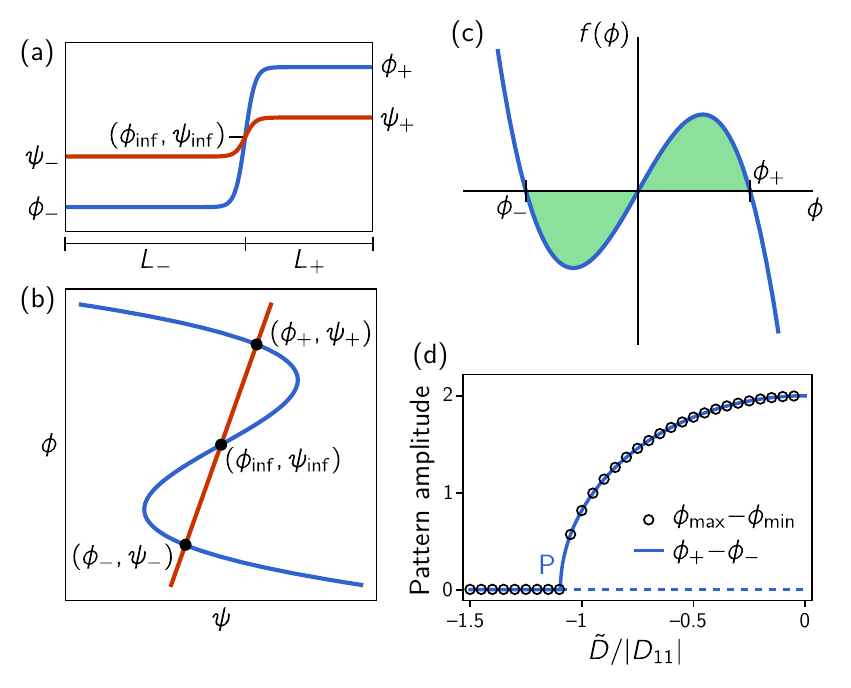}
    \caption{Graphical construction of stationary states.
    (a)~The elementary building block of a stationary pattern is a single interface separating the low density phase from the high density phase, i.e.\ a ``half-droplet''.
    (b)~$\phi$-$\psi$ phase portrait with the $\phi$-nullcline (blue line) and $\psi$-nullcline (red line), corresponding to lines where the respective generalized chemical potentials are constant [Eqs.~\eqref{eq:stat}]. The bulk phase densities and the densities at the interface's inflection point can be read of from the intersection points of the nullclines.
    (c)~Maxwell construction determining the generalized chemical potential. The areas under the curve have to balance, see Eq.~\eqref{eq:Maxwell}.
    (d)~Amplitude of stationary patterns as a function of the anti-reciprocal coupling strength $\tilde{D} < 0$. The numerical results ($\circ$) agree excellently with the prediction from the graphical construction (solid line). P marks the supercritical pitchfork bifurcation. The dashed line indicates the unstable homogeneous steady state.
    (Parameters: $\bar{\phi} = 0, D_{22} = 1.2$, $L = 100$).
    }
    \label{fig:stat-state-construction}
\end{figure}

In a stationary state $[\hat{\phi}(x), \hat{\psi}(x)]$, the fluxes of $\phi$ and $\psi$ have to vanish.
We can express these fluxes $J_\phi = -\nabla \mu_\phi$, $J_\psi = -\nabla \mu_\psi$ as gradients of the generalized chemical potentials
\begin{subequations} \label{eq:stat}
\begin{align}
    \mu_\phi(\phi, \psi) &= D_{11} \phi + \phi^3 + \tilde{D} \psi - \kappa \partial_x^2 \phi \,, \label{eq:mu-phi}\\
    \mu_\psi(\phi, \psi) &= |\tilde{D}| \phi + D_{22} \psi \,, \label{eq:mu-psi}
\end{align}
\end{subequations}
On a domain with periodic or no-flux boundary conditions, the fluxes vanish only when these potentials are spatially constant, i.e.\ $\mu_\phi(\hat{\phi}, \hat{\psi}) = \text{const}, \mu_\psi(\hat{\phi}, \hat{\psi}) \overset{!}{=} \text{const}$.
Graphically, the latter condition implies that the densities fall onto a straight line in the $\phi$-$\psi$ phase portrait, see Fig.~\ref{fig:stat-state-construction}(b). In analogy to the FHN model, we call this ``$\psi$-nullcline''. Furthermore, bulk regions where $\partial_x \phi$ vanishes must lie on the $\phi$-nullcline $D_{11} \hat{\phi} + \hat{\phi}^3 + \tilde{D} \hat{\psi} = \mu_\phi$. 
In other words, the bulk densities $\phi_{\pm}$ are given by the nullcline intersection points. The third intersection point in the center corresponds to the inflection point of the interface, where $\partial_x^2 \phi$ vanishes.

To find the generalized chemical potentials $\mu_\phi$ and $\mu_\psi$, and the size of the low and high density domains, we employ the Maxwell construction (also know as common tangent construction) and the given average densities $\bar{\phi}$ and $\bar{\psi}$.
Solving Eq.~\eqref{eq:mu-psi} for $\hat{\psi}$ and substituting into Eq.~\eqref{eq:mu-phi} gives
\begin{equation} \label{eq:stat-phi-eff}
    \underbrace{\mu_\phi - \frac{\tilde{D}}{D_{22}} \mu_\psi}_{\displaystyle \vphantom{\hat{D}_{11}} \quad := \mu = \text{const.}} = \underbrace{\left(D_{11} - \frac{\tilde{D} |\tilde{D}|}{D_{22}} \right)}_{\displaystyle \qquad := \hat{D}_{11}} \hat{\phi} + \hat{\phi}^3 - \kappa \nabla^2 \hat{\phi}\,.
\end{equation}
This is simply the stationary state equation for the Cahn--Hilliard equation. Non-trivial solutions, i.e.\ stationary patterns only exist when $\hat{D}_{11} < 0$. This can be seen from a simple mechanical analogy where the stationary profile $\hat{\phi}(x)$ is mapped to the position $\hat{\phi}$ of a ball with mass $\kappa$ at time $x$ rolling in a potential $V(\hat{\phi}) = \mu \hat{\phi} - \hat{D}_{11}\phi^2/2 + \phi^4/4$ \cite{Mikhailov1990}. Stationary patterns, corresponding to periodic oscillations of the ball exist only when $V$ has a local minimum which requires $\hat{D}_{11} < 0$.
In 1D, the value of $\mu$ is fixed as follows: We first multiply with $\partial_x \hat{\phi}$ and then integrate across the interface using $2\partial_x \partial_x^2 \hat{\phi} = \partial_x(\partial_x \hat{\phi})^2$ and the fact that $\partial_x \hat{\phi} \approx 0$ in the bulk phases, giving
\begin{equation} \label{eq:Maxwell}
    [\phi_+(\mu) - \phi_-(\mu)] \, \mu =  
    F[\phi_+(\mu)] - F[\phi_-(\mu)] \, ,
\end{equation}
with the ``local free energy density'' $F(\phi) = \hat{D}_{11}\phi^2/2 + \phi^4/4$.
For the symmetric free energy density of the Cahn--Hilliard model, Eq.~\eqref{eq:Maxwell} is solved by $\mu = 0$, $\phi_\pm = \pm \sqrt{-\hat{D}_{11}}$; see Fig.~\ref{fig:stat-state-construction}(c). The graphical construction generalizes to asymmetric local free energy densities where finding $\mu$ in general requires solving Eq.~\eqref{eq:Maxwell} numerically.
In 1D, the sizes of the low and high density domains, $L_\pm$, are determined by the average density via the ``lever rule'' $(L_+ + L_-) \, \bar{\phi} = \phi_- L_- + \phi_+ L_+$; see Fig.~\ref{fig:stat-state-construction}(a).
The above construction generalizes to higher spatial dimensions, where $x$ is taken to be the coordinate transversal to the interface. Interface curvature introduces an additional term proportional to $\kappa$ which is obtained by locally transforming to polar coordinates centered around the local center of curvature.

In the above graphical construction, the binodal lines limiting the regime of stationary patterns can be read off from the nullcline intersection in the phase portrait $\bar{\phi}_\mathrm{binodal}^\pm = \phi_\pm = \pm \sqrt{D_{11} - \tilde{D} |\tilde{D}|/ D_{22}}$. 
At $\bar{\phi} = 0$, the binodal touches the spinodal obtained from linear stability analysis, implying that the pitchfork bifurcation from the homogeneous to the phase-separated state is supercritical in that case; see Fig.~\ref{fig:stat-state-construction}(d). 
For $\bar{\phi} \neq 0$, the regime of linear instability is smaller than the regime where stationary patterns exist, implying that the system is subcritical; see Appendix~\ref{app:unequal-mix}.
This is a consequence of the fact that the spinodal and binodal become different for unequal mixtures, while they coincide for equal mixtures.
Numerical simulations show that there is also a regime of subcritical traveling waves when $\bar{\phi} \neq 0$ that emerge in a drift-pitchfork bifurcation from subcritical stationary patterns.

\subsection{Interface mode predicts traveling wave onset and speed}
\label{sec:interface-mode}

\begin{figure*}
    \centering
    \includegraphics{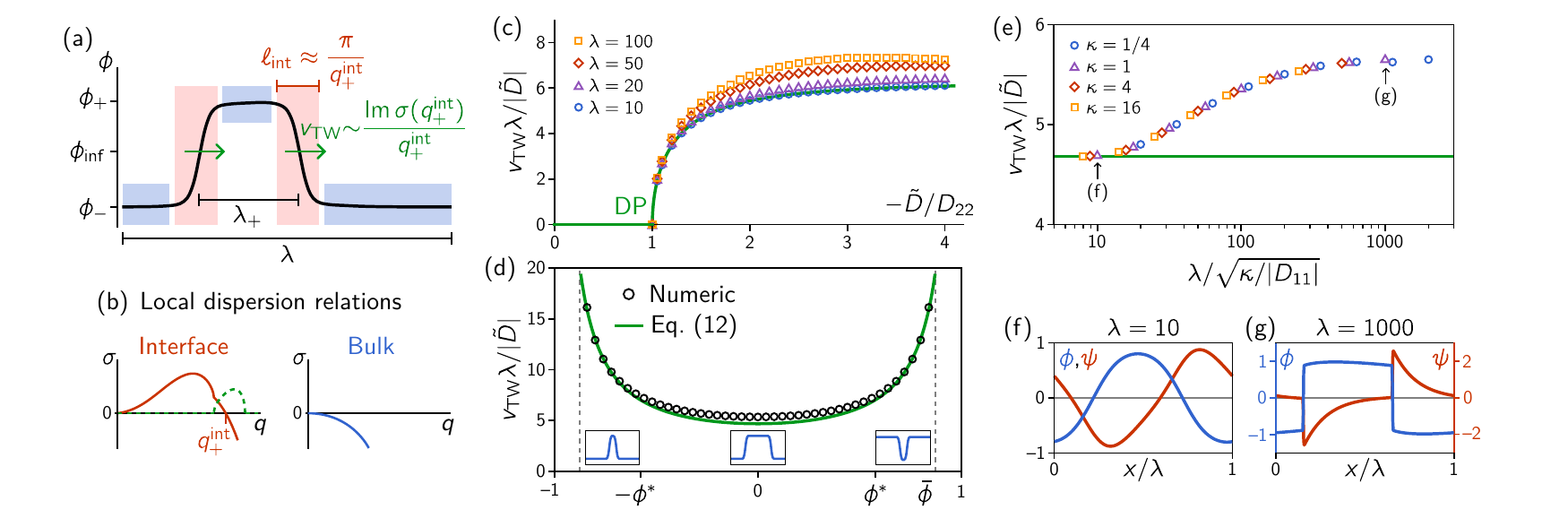}
    \caption{
    (a)~Illustration of a traveling ``droplet'' profile with the interfaces (red) and bulk (blue) regions highlighted.
    (b)~Local dispersion relations of the interface and the bulk obtained by evaluating the Jacobian Eq.~\eqref{eq:Jacobian} at $\phi_\mathrm{inf}$ and $\phi_\pm$, respectively.
    The regional dispersion relation in the bulk exhibits no instability.
    The marginal mode $q_+^\mathrm{int}$, obtained from the local dispersion relation at the interface, predicts interface width (see Ref.~\cite{Brauns.etal2020} and droplet speed as shown in (c) and (d).
    (c)~Droplet speed as a function of the nonreciprocal coupling strength $\tilde{D}$, comparing numerical simulations for various wavelengths (symbols) and the analytic prediction Eq.~\eqref{eq:wave-speed-prediction} (solid line). DP marks the drift-pitchfork bifurcation at $\tilde{D} = -D_{22}$.
    (d)~Wave speed as a function of the average density $\bar{\phi}$ which controls the relative size of the low and high density bulk domains.
    Note that the prediction from the interface mode $q_+^\mathrm{int}$ is valid even where the homogeneous steady state is unstable i.e.\ outside the ``spinodals'' $\pm\phi^* = \pm \sqrt{-D_{11}/3}$).
    No patterns exist outside the ``binodals'' indicated by vertical dashed lines.
    (e)~Wave speed as a function of the wavelength $\lambda$ and the interface width ($\ell_\mathrm{int} \approx \sqrt{\kappa/|D_{11}|}$) controlled by $\kappa$.
    (f)~For $\lambda \approx \ell_\mathrm{int}$, the wave profile is nearly sinusoidal; (g)~sharp interfaces form when $\lambda \gg \ell_\mathrm{int}$.
    [Parameters: $D_{11} = -1$, $D_{22} = 0.1$; $\bar{\phi} = 0$ in (c) and (e); $\lambda = 100$ in (d); $\tilde{D} = -0.15$ in (d) and (e).]
    }
    \label{fig:wave-speed}
\end{figure*}

The ``generalized Maxwell'' construction presented in the previous section yields the ``bulk phases'' $\phi_\pm$, but in itself is not informative about the interface that separates them.
However, from the inner nullcline intersection we can read off the profile's inflection point $(\phi_\mathrm{inf},\psi_\mathrm{inf}) = (0,0)$ around which the interface is centered [cf.\ Fig.~\ref{fig:stat-state-construction}].
This allows one to calculate a \emph{local} dispersion relation at the interface by evaluating the Jacobian Eq.~\eqref{eq:Jacobian} at $\phi_0 = \phi_\mathrm{inf}$ [see Fig.~\ref{fig:wave-speed}(b)].
This is of course a slight abuse of linear stability analysis of the homogeneous state since the interface is evidently not homogeneous. Nonetheless, the local dispersion relation is informative about the instability that maintains the interface against ``flattening'' towards a homogeneous state. The interface width $\ell_\mathrm{int}$ is the length scale where destabilizing and stabilizing effects balance and can be read off from the marginally stable mode $q_+^\mathrm{int}$ in the dispersion relation: $\ell_\mathrm{int} \approx \pi/q_+^\mathrm{int}$ \cite{Brauns.etal2020}.
More precisely, the interface can be approximated as a half-period sinusoidal profile $\phi \sim \sin[q_+^\mathrm{int} (x - x_\mathrm{int})]$ concatenated to flat plateaus at $\phi = \phi_\pm$ [see Fig.~\ref{fig:wave-speed}(a)]. 
To understand how the interface width is selected dynamically, consider an interface wider than $\pi/q_+^\mathrm{int}$. The corresponding sinusoidal mode will grow, since it falls into the band of unstable modes ($q < q_+^\mathrm{int}$). The nonlinearity causes the saturation of $\phi$ at the plateaus $\phi_\pm$, while the sinusoidal mode's growth causes the interface to narrow. This process stops when the interface width reaches $\pi/q_+$ where the sinusoidal mode at the interface stops growing.
Conversely, an interface initially narrower than $\pi/q_+^\mathrm{int}$, will contain a sinusoidal mode that decays until the interface width reaches $\pi/q_+^\mathrm{int}$.

For sufficiently strong anti-reciprocal coupling ($\tilde{D} < -D_{22}$), the interface mode's growth rate acquires a non-zero imaginary part [$\Im \sigma(q^\mathrm{int}_+) \neq 0$, see Fig.~\ref{fig:LSA}(a)], indicating a propagating mode driven by local chase-and-run dynamics.
Indeed, this condition coincides exactly with the drift-pitchfork bifurcation where stationary patterns turn into traveling waves [cf.\ Fig.~\ref{fig:LSA}(g)]. 
What sets the speed of these waves?
The propagation speed of a mode with wavenumber $q$ and complex growth rate $\sigma(q)$ is given by $\Im \sigma(q)/q$. Based on the hypothesis that the waves are driven by the interface mode, we expect that the speed of traveling waves follows the relationship 
\begin{equation} \label{eq:interface-drift}
    v_\mathrm{TW} \propto \frac{\Im \sigma(q^\mathrm{int}_+)}{q^\mathrm{int}_+}.
\end{equation}
In fact, this relationship holds as an equality for a sinusoidal pattern where the wavelength is twice the interface width \cite{You.etal2020}.
Since the droplet propagation requires redistribution of mass, the wave speed is limited by diffusive transport through the domains separating the interfaces when $\lambda \gg \ell_\mathrm{int}$.
Specifically, the mass fluxes through these domains scales as $J_\pm \propto 1/\lambda_\pm$, where $\lambda_-$ ($\lambda_+$) denote the widths of the low (high) density domains.
Moving an interface bridging the density difference $\phi_+ - \phi_-$ requires a net mass flux
\begin{equation} \label{eq:wave-speed_vs_lengths}
    (\phi_+ - \phi_-) v_\mathrm{TW} = J_+ + J_- \propto \frac{1}{\lambda_+} + \frac{1}{\lambda_-} \, .
\end{equation}
To fix the prefactor, we demand that the relation $v_\mathrm{TW} = \Im \sigma (q_+)/q_+$ is recovered for the sinusoidal pattern where $\lambda_+ = \lambda_- = L/2 \approx \pi/q_+$. This yields
\begin{align}
    v_\mathrm{TW} &\approx \frac{\pi \Im \sigma(q^\mathrm{int}_+)}{2 \, (q^\mathrm{int}_+)^2} \left(\frac{1}{\lambda_+} + \frac{1}{\lambda_-} \right),\\
    &= \frac{|\tilde{D}|}{\lambda} \, \frac{2 \pi}{1 - \bar{\phi}^2} \, \sqrt{1 - \frac{D_{22}^2}{\tilde{D}_{\vphantom{2}}^2}} \, . \label{eq:wave-speed-prediction}
\end{align}
In the second line, we used Eq.~\eqref{eq:Im} and the ``lever rule'' for the sizes of the low and high density domains. Notably, $q^\mathrm{int}_+$ cancels from this expression implying that $v_\mathrm{TW}$ is independent of the interface width.
In numerical simulations across parameter sweeps for $\tilde{D}$, $\bar{\phi}$, $\lambda$ and $\kappa$, we find that the observed wave speeds agree well with the analytic approximation from Eq.~\eqref{eq:wave-speed-prediction} as shown in Fig.~\ref{fig:wave-speed}(c--e).
Remarkably, the prediction is accurate even in the binodal regime $|\bar{\phi}| > \phi^* = \sqrt{-D_{11}/3}$, where the homogeneous steady state is stable.
This emphasizes the fact that the \emph{local} dispersion relation at the interface informs about properties of the highly nonlinear pattern which are not captured by the global dispersion relation of the homogeneous steady state.

In passing, we note independence of wave speed from the interface width suggests that one might derive an analytic expression in the sharp interface limit $q^\mathrm{int}_+ \to \infty$. 
However, the cross-diffusive coupling poses a particular challenge for such an approach since $\psi$ becomes discontinuous at interfaces in the limit. This problem is not present in recently studied systems, where a phase separating field is coupled to a non-conserved diffusive field via source terms rather than cross-diffusive fluxes such that a sharp interface approximation can be applied straightforwardly \cite{Alert.etal2022,Zhao.etal2023,Demarchi.etal2023,Goychuk.etal2024}. Resolving this technical challenge is left for future work.

\section{Interrupted coarsening and wavelength selection}
\label{sec:coarse}

\begin{figure}
    \centering
    \includegraphics{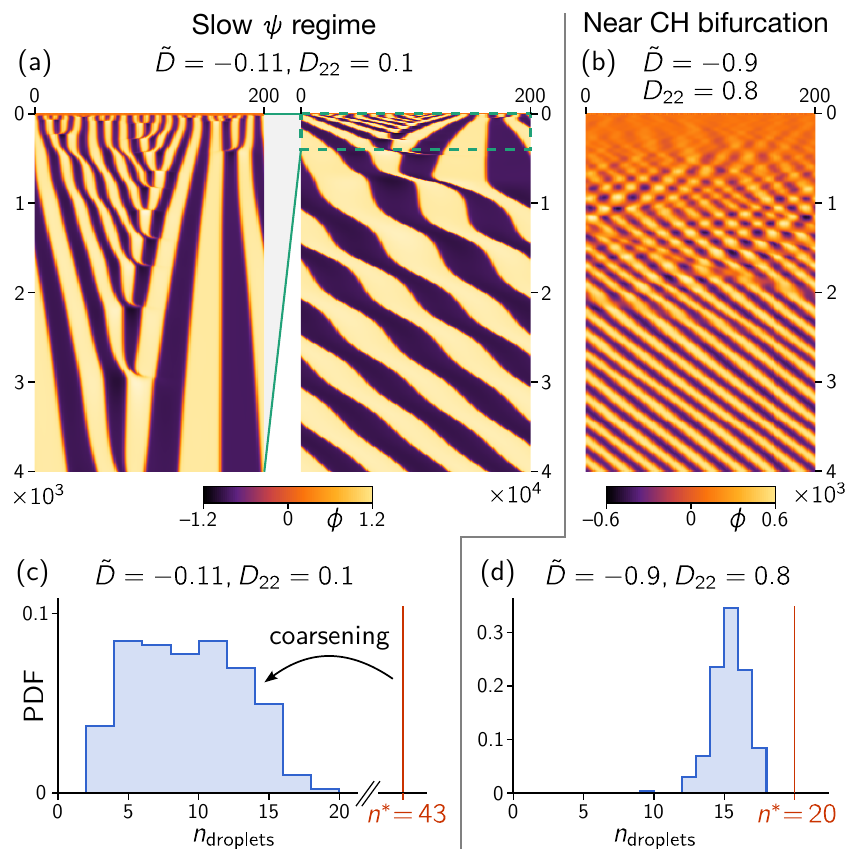}
    \caption{
    (a)~In the slow-$\psi$ regime, droplet collisions drive coarsening (see blow up view on the left) until all droplets travel in the same direction and slowly equilibrate their masses (right).
    (b)~Near the CH bifurcation, the dynamics appears as a superposition of counter-propagating traveling waves, corresponding to the fastest growing mode of the dispersion relation. Over time, the amplitude of one of the waves vanishes and the steady state is always a pure traveling wave.
    (c),~(d)~Distribution of droplet number (inversely proportional to wavelength) of traveling wave trains in simulations initialized from a homogeneous state with small amplitude noise. Red lines indicate the number of ``droplets'' corresponding to the fastest growing mode, $n^* = L q_\mathrm{max}/2\pi$.
    (c)~Significant coarsening and broad distribution of final peak number near the drift-pitchfork bifurcation.
    (d)~Almost no coarsening and narrow distribution of final peak number near the supercritical conserved-Hopf bifurcation.
    [Parameters: $L = 200$  in (b) and (c) and $L=400$ in (c) and (d); ensemble size 200. Simulations were run sufficiently long to ensure that a steady state was reached, $T = 5{\times}10^5$.]
    }
    \label{fig:wavelength-selection}
\end{figure}

A hallmark of phase-separating systems is coarsening, the growth of the characteristic length scale of spatial patterns.
In mass-conserving systems, coarsening is driven by redistribution of mass from smaller to larger droplets, a process known as Ostwald ripening \cite{Lifshitz.Slyozov1961,Wagner1961}.

In the reciprocally coupled case, coarsening in the NRCH system proceeds to complete phase separation as in the conventional Cahn--Hilliard equation [cf.\ Fig.~\ref{fig:FHN}(c)].
We find the same behavior for weak anti-reciprocal coupling ($-\tilde{D} < D_{22}$), where no traveling waves occur.
By contrast, the coarsening kinetics is markedly different in the region where the system supports  traveling waves ($-\tilde{D} > D_{22}$).

In the slow-$\psi$ regime ($|\tilde{D}| \ll |D_{11}|$), traveling droplets initially coarsen via collisions of counter-propagating droplets as shown in the kymograph in Fig.~\ref{fig:wavelength-selection}(a). 
Remarkably, the coarsening stops once all droplets travel in the same direction. 
Mass is redistributed from smaller to larger droplets until their masses are equilibrated, resulting in a periodic wavetrain. 
Starting from a homogeneous state perturbed by small amplitude noise, the initial propagation directions of the droplets are random. As a result, an ensemble of simulations with different noise realizations shows a broad distribution of droplets remaining in steady state [Fig.~\ref{fig:wavelength-selection}(c)].
In other words, the dynamics does not select a particular wavelength. 
In fact, starting simulations initialized with a given wavelength, we find that the initial wavelength is selected in the final steady state, implying that all wavelengths (larger than the interface width) are stable. In particular, this includes the fully phase separated state (see Fig.~\ref{fig:TW-multistability} in Appendix~\ref{app:numerics}).

We find this remarkable multistability of wavelengths for all traveling waves.
However, close to the CH bifurcation, i.e.\ in the fast-$\psi$ regime $|\tilde{D}| \approx |D_{11}|$,
we find that the final wavelength selected from a randomly perturbed initial condition
is very close to the wavelength of the fastest growing mode [see Fig.~\ref{fig:wavelength-selection}(b) and Fig.~\ref{fig:wavelength-selection}(d)].
This suggests that a different wavelength selection mechanism is at play in this regime.
Recall that the CH bifurcation is supercritical, meaning that the pattern amplitude is small in the vicinity of the bifurcation and the dynamics is dominated by the fastest growing mode. Since this mode is propagating near the CH bifurcation [cf.\ Fig.~\ref{fig:LSA}(d)], there is initially a superposition of two counter-propagating waves [Fig.~\ref{fig:wavelength-selection}(b)]. These waves interact through the nonlinearity, which eventually causes the amplitude of one to go to zero such that a traveling wave remains. 
Alternatively, the amplitudes of both waves might equilibrate, which would result in a standing wave pattern. However, we have not found stable standing waves in any of our simulations, suggesting that they are unstable. An analysis in terms of the amplitude equation formalism (also known as weakly nonlinear analysis) \cite{Cross.Hohenberg1993} might explain this puzzling observation.

The multistability of wavetrains with different wavelengths, including the fully phase separated state, is similar to what is found in 1D models of flocking \cite{Caussin.etal2014,Solon.etal2015}. However, in that case not all wavelength larger than the interface width are stable, and in particular short wavelengths are unstable. 
Other scenarios for interrupted coarsening, such as in systems with weakly broken mass conservation, long range interactions, or coupling to chemical interactions generally exhibit a band of stable wavelengths that is bounded both from above and below \cite{Brauns.etal2021,Zwicker2022}. The reason for this is that the same process that causes coarsening to stop also drives splitting of domains above a critical size \cite{Zwicker.etal2017,Brauns.etal2021}.

\begin{figure*}[tb]
    \centering
    \includegraphics{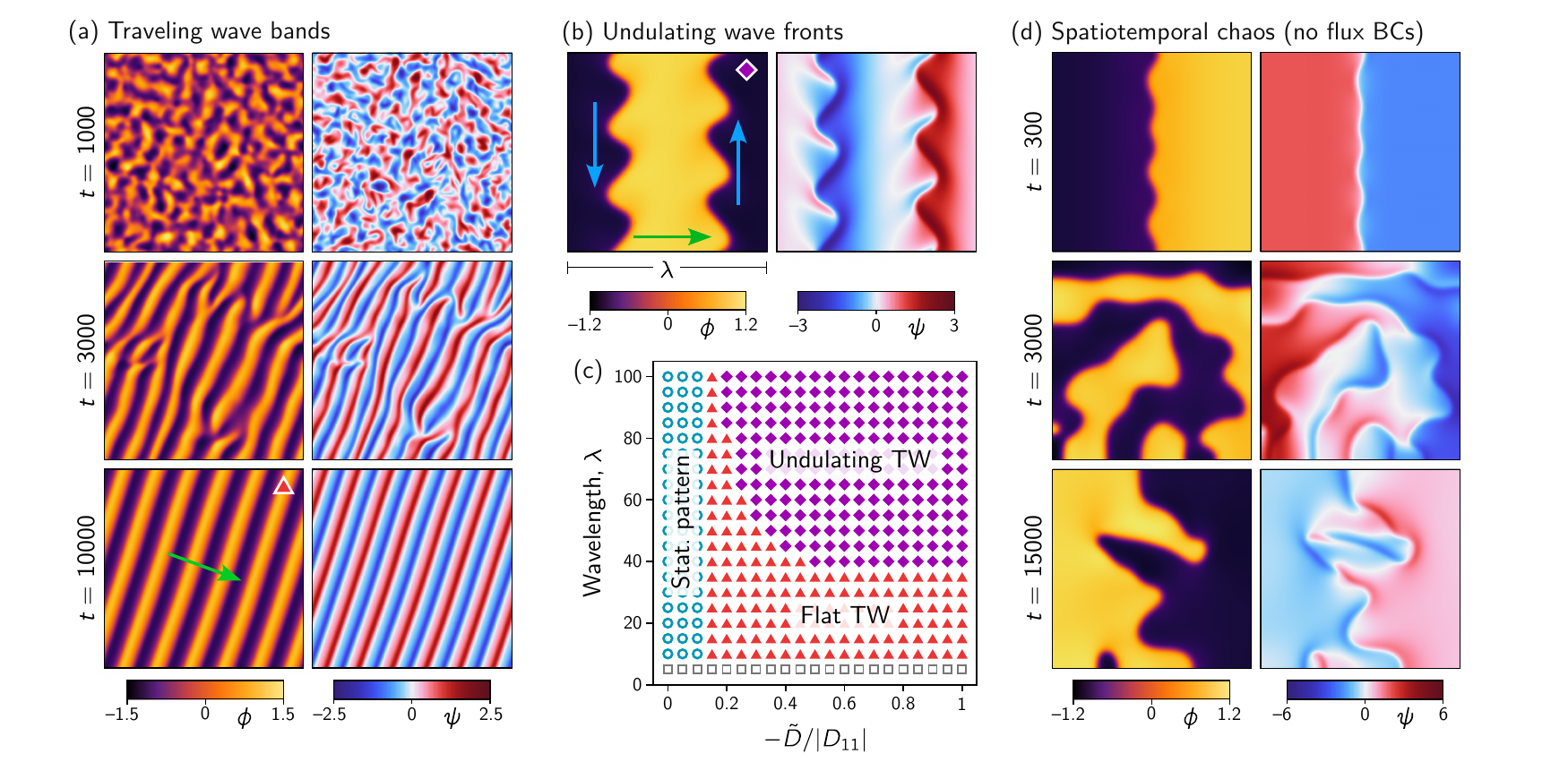}
    \caption{
    (a)~Traveling wave bands form in a 2D domain with periodic boundary conditions (see Movie~1; $\tilde{D} = -1$). The green arrow indicates the propagation direction.
    (b)~Undulations form spontaneously along the interfaces of traveling waves (see Movie~2 and Movie~3). The undulations travel along the interfaces  (blue arrows), i.e.\ transversal to the traveling wave direction (red arrow).
    (c)~Phase diagram showing that transversal undulations (purple diamonds) form only for traveling waves whose wavelength $\lambda$ is sufficiently large. The interfaces of traveling waves with smaller wavelength remain flat. Simulations were performed in a square domain with side length $\lambda$ and initialized with a single high-density band, corresponding to a traveling wave with wavelength $\lambda$.
    (d)~Undulations destabilize interfaces and thereby drive spatiotemporal chaos in simulations with no-flux boundary conditions ($\tilde{D} = -0.2$); see Movie~4.
    [$\bar{\phi} = 0, D_{22} = 0.1$; $L = 200$ in (a); $L = 100$ in (b) and (d).]
    }
    \label{fig:2D-patterns}
\end{figure*}

\section{2D patterns: undulating fronts and spatiotemporal chaos}
\label{sec:2D}

We conclude our analysis of Eqs.~\eqref{eq:NRCH} with a brief exploration of dynamics in two spatial dimensions. 
As in one dimension (1D), we find uninterrupted coarsening in the phase separation regime and interrupted coarsening for traveling waves [see Fig.~\ref{fig:2D-patterns}(a) Movie~1]. 
Strikingly, for sufficiently large wavelength, the interfaces of the traveling waves start undulating and the undulations propagate transversal to the traveling wave [see Fig.~\ref{fig:2D-patterns}(b) and Movies~2 and~3]. 
While traveling waves break polar symmetry, the traveling undulations additionally break chiral symmetry. 
Notably, undulations of adjacent interfaces can travel either in the same direction (see Movie~2) or in opposite directions [see Fig.~\ref{fig:2D-patterns}(b) and Movie~3].

A large parameter sweep shows that the threshold wavelength of traveling bands above which undulations form decreases with the strength of nonreciprocal coupling [Fig.~\ref{fig:2D-patterns}(c)]. 
We suspect that the interface undulations are driven by a nonreciprocal generalization of the Mullins--Sekerka instability \cite{Mullins.Sekerka1963} which describes fingering of a propagating interface and is driven by gradients of a single diffusive field. 
The nonreciprocal interaction with the second diffusive field is likely responsible for the transversal motion of the undulations, which is absent in the classical Mullins--Sekerka fingers.
A further investigation of the interface instability using capillary wave theory is beyond the scope of this manuscript and deferred to future work.
In the Conclusion (Sec.~\ref{sec:discussion}), we briefly discuss potential relations to other interfacial instabilities discovered in recent studies \cite{Saha.etal2020,Fausti.etal2021,Alert.etal2022}.

For non-equal mixtures ($\bar{\phi} \neq 0$), the canonical Cahn--Hilliard dynamics transitions from labyrinth-like patterns to high density droplets in a low density background ($\bar{\phi} < 0$). [Since the model is symmetric under $\phi \to -\phi$, the dynamics of droplets is equivalent to that of ``bubbles/holes''.]
The NRCH model shows a similar transition from traveling wave bands to traveling droplets \cite{Saha.etal2020} that can form regular dynamic lattices (see Movie~6). In large systems, we find that interfacial undulations cause larger droplets to break up while smaller droplets collide and merge resulting in chaotic dynamics (see Movie~7). 
The breaking up of large droplets interrupts coarsening and the system eventually reaches a steady distribution of droplet sizes (see Fig.~\ref{fig:droplet-size-distribution} in Appendix~\ref{app:numerics}).
This dynamically maintained distribution of droplet sizes is in contrast to the stable traveling wave (``train of droplets'') which forms in 1D because droplet collisions cease once all droplets travel in the same direction (cf.\ Fig.~\ref{fig:wavelength-selection}).
As a consequence, we do not expect that interrupted coarsening in 2D will exhibit the same strong dependence on the initial condition as we found in 1D.
Testing this hypothesis and developing a kinetic theory for the droplet size distribution are interesting avenues for future research.

Finally, let us turn to 2D systems with no-flux boundary conditions. In contrast to 1D, where no-flux boundaries can arrest the interface mode that drives traveling waves (see Appendix~\ref{app:no-flux}), the propagating interface mode can drive mass transport transversal to the interface in 2D. 
Thus, stationary interfaces are always unstable in 2D for $\tilde{D} < -D_{22}$ [Fig.~\ref{fig:2D-patterns}(d)].
Starting from a fully phase separated (demixed) initial condition, the interfaces' transversal instability drives the breakup of the phase-separated domains, leading to a dynamic, ``microphase-separated'' state, where mass sloshes from one side of the domain to the other chaotically [see Fig.~\ref{fig:2D-patterns}(d) and Movie~4]. 
The slow sloshing is a result of the initial condition where $\phi$ is concentrated on one side of the system. Starting from a homogeneous state, we find spatiotemporal chaos (see Movie~5). 

\section{Concrete physical systems} 
\label{sec:systems}

In the following, we discuss various classes of systems that can be mapped to Eqs.~\eqref{eq:NRCH} as summarized in Table~\ref{tab:models}.
The appearance of propagating modes near $q_+$ is the characteristic feature that identifies these systems, since it indicates that their linearized dynamics (Jacobian) is of the form Eq.~\eqref{eq:Jacobian}.
For each system, we discuss the origin of the effective negative diffusion $D_{11}$ that drives the pattern-forming (mass-redistribution) instability and the origin of nonreciprocity in the effective cross-diffusive coupling.

Before discussing concrete physical systems, we note that mapping to Eqs.~\eqref{eq:NRCH}, will generally lead to coefficients $D_{ij}$ and $\kappa$ that depend on the fields $\phi$ and $\psi$. 
Nonetheless, studying the minimal model provides a crucial baseline. The rich behavior of the NRCH model suggests that in many cases the minimal model might indeed be sufficient to provide a phenomenological account of experimental observations. The failure of the minimal model to reproduce certain experimental observations may provide hints toward necessary model extensions, such as non-constant coefficients~\cite{Saha.Golestanian2022}.

In \emph{nonreciprocally coupled binary mixtures} \cite{Saha.etal2020,Saha.Golestanian2022,Frohoff-Hulsmann.etal2021}, phase separation is driven by relaxation to equilibrium while nonreciprocity results explicitly from non-equilibrium effects. This is in contrast to the other systems below, in which phase separation itself is driven by non-equilibrium processes such that there is no equilibrium limit which is reciprocal but still exhibits phase separation. 

For \emph{mixtures of active and passive Brownian particles}, a mapping to Eqs.~\eqref{eq:NRCH} has been provided in Ref.~\cite{You.etal2020}.
The active particles form patterns via motility-induced phase separation \cite{Cates.Tailleur2015}. Persistence of self-propulsion of the active particles causes them to slow down when their orientation points in a direction of increasing density (of both active and passive particles). This causes both the mass-redistribution instability and the negative cross-diffusion term $\chi_\mathrm{ap}$ (here $D_{12}$). In contrast, the cross-diffusion term $\chi_\mathrm{pa}$ (here $D_{21}$) is not affected by activity, because passive particles are simply sterically repelled by both active and passive particles.

\emph{Mass-conserving reaction--diffusion systems} describe, for instance, pattern formation by proteins that cycle between different conformational states (e.g.\ membrane bound and cytosolic) much faster than they are produced or degraded \cite{Halatek.etal2018}.
An effective description of the proteins' mass-redistribution dynamics can be obtained via a local equilibrium approximation \cite{Halatek.Frey2018,Brauns.etal2020,Brauns.etal2021a}.
The most widely studied example is the Min-protein system of \textit{E.\ coli}, which has been reconstituted in vitro \cite{Loose.etal2008}, where it exhibits a remarkable diversity of spatiotemporal patterns \cite{Ivanov.Mizuuchi2010,Glock.etal2019,Brauns.etal2021b}. 
The dynamics of the densities of MinD and MinE can be mapped to Eq.~\eqref{eq:NRCH} via the local equilibrium approximation (see Appendix~\ref{app:MCRD} for details).
Due to MinD's self-recruitment to the membrane, its effective self-diffusion coefficient is negative thus driving the pattern forming mass-redistribution instability \cite{Brauns.etal2021a}. Nonreciprocal cross-diffusion of MinD and MinE is a consequence of the nonreciprocity in the chemical reactions (MinD recruits MinE to the membrane while MinE drives membrane detachment of MinD).
In general, the effective transport coefficients can be read off from the slopes of the surfaces of reactive equilibria as a function of the total densities, which gives them a simple geometric interpretation~\cite{Brauns.etal2021a}. 
We note that a systematic mapping of mass-conserving reaction--diffusion dynamics to a more general form of the nonreciprocal Cahn--Hilliard equation has been obtained very recently via a weakly nonlinear analysis in Ref.~\cite{Frohoff-Hulsmann.Thiele2023}. While this approach is mathematically more rigorous than ours, it does not provide the physical and geometric insight afforded by the local-equilibrium approximation.

\emph{Active gel theory} has been used to model active viscoelastic and poroelastic media on various scales from the intracellular actomyosin cortex \cite{Bois.etal2011,Radszuweit.etal2013,Banerjee.etal2017} to tissues and motile cells embedded in extracellular matrix (ECM) \cite{Oster.etal1983,Palmquist.etal2022, Yin.Mahadevan2022}. These models qualitatively reproduce the wave-like and ``pulsatory'' dynamics observed in these biological systems \cite{Bement.etal2015,Michaux.etal2018,Wigbers.etal2021,Serra-Picamal.etal2012}.
A prototypical model for motile cells embedded in ECM serves as a concrete setting in the following discussion (mapping of poroelastic models \cite{Radszuweit.etal2013,Weber.etal2018} to these equations is briefly discussed in Appendix~\ref{app:poroelastic}).
In the following, we restrict our analysis to \emph{isotropic} active stress and a 1D description that captures only the longitudinal mode, where the displacement field is aligned with the wavevector. This setting is sufficient to capture the basic instability of active contractile media \cite{Oster.etal1983,Bois.etal2011,Radszuweit.etal2013,Weber.etal2018}.
In future work, it will be interesting to study 2D and 3D active gel models, where the transversal component(s) of the displacement field may be excited by the undulational instability of traveling waves.

\begin{table*}
    \caption{
    Overview of physical systems sharing the same prototypical core behavior described by Eq.~\eqref{eq:NRCH}.
    The Min-protein system of \textit{E.\ coli} serves as example for protein-based pattern formation described by mass-conserving reaction diffusion systems.  
    \textsuperscript{*}MinE redistribution is required for the pattern-forming instability, see Ref.~\cite{Brauns.etal2021a}.
    \textsuperscript{$\dagger$}The density field alone does not possess an instability in this case: coupling to strain is required for the instability, see Eq.~\eqref{eq:active-gel-effective}.
    \textsuperscript{$\ddagger$}Hydrodynamic mode due to translation invariance of the displacement field $u \to u + a$.
    }
    \renewcommand{\arraystretch}{1.3}
    \begin{tabular*}{\linewidth}{@{}l@{\extracolsep{\fill}}ll@{}}
        \toprule
        System &  $\phi$ &  $\psi$ \\
        \midrule
        Nonreciprocal binary mixtures \cite{You.etal2020, Curatolo.etal2020,Saha.etal2020,Saha.Golestanian2022,Dinelli.etal2023,Duan.etal2023} &  
        Phase-separating field
        & Diffusive field \\
        Active/passive particle mixtures \cite{You.etal2020} & Density of active particles & Density of passive particles \\
        Mass-conserving reaction--diffusion systems \cite{Brauns.etal2021b,John.Bar2005a,Jacobs.etal2019} & MinD concentration\textsuperscript{*} & MinE concentration \\
        Active gels \cite{Oster.etal1983, Weber.etal2018, Bois.etal2011, Radszuweit.etal2013, Yin.Mahadevan2022} & Density of contractile elements\textsuperscript{$\dagger$} & Strain\textsuperscript{$\ddagger$} \\
        \bottomrule
    \end{tabular*}
    \label{tab:models}
\end{table*}

In 1D, mass-conservation of cell density $c$ and force balance take the form
\begin{subequations} \label{eq:active-gel}
\begin{align}
    \partial_t c + \partial_x(\dot{u} c) &= D \partial_x^2 c \label{eq:cell-density}\\
    \gamma \dot{u} &= \partial_x \big[ \eta \partial_x \dot{u} + E \partial_x u + T_\mathrm{a}(c) \big]\,.
    \label{eq:force-balance}
\end{align}
\end{subequations}
Here $u$ is the displacement (and $\dot{u} = \partial_t u$ the velocity) of the ECM which is modelled as a Kelvin--Voigt viscoelastic material with stiffness $E$ and viscosity $\eta$, $\gamma$ is the friction of the ECM with an underlying rigid substrate, and $T_\mathrm{a}(c)$ is the active stress exerted by the cells. 

Let us first consider the case of vanishing ECM stiffness $E = 0$. This case corresponds to the model studied in Ref.~\cite{Bois.etal2011} in the context of an actomyosin cortex, where $c$ describes the density of contractile myosin motors.
We can formally solve Eq.~\eqref{eq:force-balance} and substitute into Eq.~\eqref{eq:cell-density}
\begin{equation} \label{eq:cell-density-effective}
    \partial_t c = \partial_x \big\{ \big[D - c \, (\gamma - \eta \partial_x^2)^{-1} T_\mathrm{a}'(c) \big] \, \partial_x c \big\}\,,
\end{equation}
where $T_\mathrm{a}' = \partial_c T_\mathrm{a}$.
This is a closed equation for $c$ and has the form of a diffusion equation where the effective diffusion constant becomes negative in the long wavelength limit when $c T_\mathrm{a}'(c) > \gamma D$.

Going through the same calculations for a finite stiffness $E > 0$ and introducing the strain $\varepsilon = \partial_x u$ yields (see Appendix~\ref{app:poroelastic})
\begin{subequations} \label{eq:active-gel-effective}
\begin{align}
    \gamma \partial_t c &= \partial_x \big[(\gamma D - c T_\mathrm{a}') \partial_x c - c E \, \partial_x \varepsilon \big] - \tfrac{\eta}{\gamma}\partial_x^4 T_a(c) \, , \\
    \gamma \partial_t \varepsilon &= E \, \partial_x^2 \varepsilon + \partial_x (T_\mathrm{a}' \, \partial_x c) \,, \label{eq:strain-effective}
\end{align}
\end{subequations}
where we have expanded the kernel $(\gamma - \eta \partial_x^2)^{-1}$ to second order in $\partial_x$ to obtain the term $\tfrac{\eta}{\gamma} \partial_x^4 T_a(c)$ that stabilizes short wavelenghts.
In the form Eq.~\eqref{eq:active-gel-effective}, it becomes clear that ECM elasticity has the role of an effective ``cross-diffusive'' coupling that couples the cell density $c$ to the ECM strain $\varepsilon$. In turn, the active tension $T_\mathrm{a}(c)$ provides the cross-coupling from $c$ to $\varepsilon$.  
Because the ``cross-diffusion'' terms appear with opposite signs (and because $T_\mathrm{a}'(c) > \gamma D/c > 0$ is required for phase separation), the coupling is necessarily anti-reciprocal. 
Traveling waves therefore appear generically. 

The hydrodynamic modes in active gels arise from mass-conservation of the density $c$ and translation invariance of the displacement field $u$ (i.e.\ invariance under $u \to u + a$).
The translation invariance is broken if the ECM is elastically attached to an underlying rigid substrate, which introduces a term $k \varepsilon$ in Eq.~\eqref{eq:strain-effective}. This is analogous to a production--degradation term that breaks mass conservation. This generically interrupts the coarsening process leading to microphase separation \cite{Brauns.etal2021}.
For soft attachment (small $k$), the branches crossing in the dispersion relation remains close to $q_+$, such that the interfacial mode remains oscillatory/propagating. Stiff attachment however shifts the second branch of the dispersion relation down and thereby suppresses traveling waves.

A slightly different scenario to the one described by Eq.~\eqref{eq:active-gel} is a viscous active gel that is coupled to an elastic substrate by friction. In this case, the substrate needs to be sufficiently soft for traveling waves to emerge. In the limit of a rigid substrate, one recovers the scenario studied in Ref.~\cite{Bois.etal2011}.

Contraction pulses and waves are observed in the actomyosin cortex of cells \cite{Michaux.etal2018,Bement.etal2015,Wigbers.etal2021,Landino.etal2021} and in tissues \cite{Serra-Picamal.etal2012,Boocock.etal2021}. While most previous models have relied on oscillatory or excitable biochemical kinetics \cite{Bement.etal2015,Staddon.etal2022,Wigbers.etal2021}, our unifying NRCH model demonstrates how these dynamic patterns can arise generically from transport of conserved quantities even in the absence of feedback loops in the biochemical reactions. A key feature of the NRCH model is the transition from dynamic to stationary patterns as a function of the (effective) transport coefficients. This might shed new light on the question how the actomyosin cortex transitions from dynamic pulses to the stationary cytokinetic ring \cite{Cao.Wang1990}.

\textit{Chemosensitive motility and catalytically active droplets.}
The cell densities of two motile bacterial species that cross-regulate each others motility via signaling molecules can effectively be described as a binary mixture with nonreciprocal coupling \cite{Curatolo.etal2020}. Non-reciprocity of the effective cross-diffusion results directly from the nonreciprocity of chemical signaling interactions.
The effective description in terms of only the bacterial densities is valid when the dynamics of the signaling molecules is much faster than the bacterial motility.

Recently, the opposite limit of a fast chemotactic particles coupled to a slow signaling molecule has been studied in Ref.~\cite{Zhao.etal2023}. Even though the chemical field is not conserved in the dynamics, the branches in the dispersion relation cross near $q_+$ when the chemical field is slow. 
Therefore, the traveling waves emerge by the same mechanism as in the prototypical model Eq.~\eqref{eq:NRCH}.
A related scenario was studied in Ref.~\cite{Demarchi.etal2023} in the context of catalytically active droplets.
In both models \cite{Zhao.etal2023,Demarchi.etal2023}, broken mass conservation leads to interrupted coarsening.
As we have shown here, coarsening of traveling waves is generically interrupted, even without broken mass conservation.
The mechanism of wavelength selection due to weakly broken mass conservation is well understood for quasi-stationary patterns \cite{Glotzer.etal1995,Brauns.etal2021,Weyer.etal2023},
In contrast, for traveling wave patterns the question of wavelength selection far from a homogeneous state remains a wide open and an interesting avenue for future research.

\section{Conclusion}
\label{sec:discussion}

Traveling waves emerge generically in a broad range of dynamical systems. While their emergence and dynamics are well understood in excitable and oscillatory media without conserved quantities \cite{Cross.Hohenberg1993,Mikhailov1990,Tyson.Keener1988,Aranson.Kramer2002}, much less is known about traveling waves in the presence of conservation laws. Here, we have presented the NRCH model Eq.~\eqref{eq:NRCH} as a minimal model for the emergence of traveling waves in systems with conservation laws.
The NRCH model can be seen as the mass-conserving analogue of the well-known FitzHugh-Nagumo reaction--diffusion model \cite{Rocsoreanu.etal2000}; see Table~\ref{tab:FHN-CH}. It unifies many previously studied systems, including mixtures of active and passive particles, reaction--diffusion systems and active gels [see Table~\ref{tab:models}].
Notably, already this minimal model exhibits several remarkable behaviors such as coarsening driven by droplet collisions which arrests without selecting a preferred wavelength and a transversal instability of planar interfaces that gives rise to traveling wavefront undulations and spatiotemporal chaos.
The present study provides only a first glimpse of this rich phenomoenology and raises many interesting questions for further research as summarized below.

As a central result, we have identified the mechanism by which stationary patterns transition to traveling patterns in the NRCH model. The transition is heralded by an exceptional point in the \emph{local} dispersion relation at the interfaces of the stationary pattern.
At the exceptional point, two hydrodynamic modes coalesce i.e.\ the Jacobian's eigenvectors coincide.
This mechanism of temporal organization generalizes the notion of nonreciprocal phase transitions \cite{Fruchart.etal2021} to mass-conserving systems.
In the non-conserved case, the mode coalescence involves a global Goldstone mode (e.g.\ due to the rotation invariance of an orientational order parameter or oscillator phase) and is independent of spatial gradients. 
By contrast, in the mass-conserving system, the mode coalescence takes place at finite wavenumber and is spatially localized to domain interfaces.
Notably, the traveling wave speed is predicted by the local dispersion relation at the interface, even far from the homogeneous steady state.

More generally, our results show that interfaces are the relevant collective excitations that govern the pattern dynamics of nonreciprocally coupled conserved fields.
This justifies, a posteriori, that we used the minimal form of the equations for two conserved fields Eqs.~\eqref{eq:NRCH}, as these equations capture the key features: the formation of interfaces through phase separation and their motion governed by the local dispersion relation.  
This offers a complementary approach to the amplitude equation formalism, which is valid for small amplitude sinusoidal patterns near the onset of pattern formation (e.g.\ near the conserved Hopf bifurcation) \cite{Frohoff-Hulsmann.Thiele2023}.

We are only beginning to understand the rich phenomenology of the NRCH model and many important questions remain open. In the following, we point to several exciting avenues for future research.

First, noise plays a crucial role in the nonreciprocal phase transitions studied in \cite{Fruchart.etal2021}. This is not the case for the systems studied here. On the other hand, our analysis of the deterministic dynamics has revealed that traveling waves are highly multistable, with stable wavelengths ranging from the interfacial length scale up to the system size (corresponding to a fully phase separated state) and the selected wavelength depends sensitively on the initial condition. 
This suggests that noise may play an important role in wavelength selection, similar to what is observed in models of flocking \cite{Solon.etal2015}.
Other than noise, weakly broken mass-conservation \cite{Glotzer.etal1995,Brauns.etal2021}, coupling to chemical reactions \cite{Zwicker2022}, or long-range interactions \cite{Liu.Goldenfeld1989,Ohta.Kawasaki1986}, e.g.\ via an elastic medium \cite{Rosowski.etal2020,Qiang.etal2023}, may also provide mechanisms for wavelength selection. 
The scenario of elasticity-mediated long-range coupling is particularly interesting in the context of active gel theories and poroelastic media.

Second, we found the emergence of transversally traveling undulations at the interfaces of traveling waves as a particularly striking feature of the NRCH dynamics.
These undulations are reminiscent of the traveling finger patterns observed at an oil-air interface driven by the rotation of two acentrically mounted cylinders~\cite{Pan.DeBruyn1994}.
A similar phenomenon has recently been observed in simulations of a nonreciprocal four-component mixture \cite{Saha.etal2020}.
While the exact mechanism driving these undulations remains unclear, we hypothesize that they may arise via a nonreciprocal generalization of the well-known Mullins--Sekerka instability \cite{Mullins.Sekerka1963}. 
An active Mullins--Sekerka instability has recently been identified in a model for active phase separation \cite{Fausti.etal2021}, where the amplitude of undulations does not saturate leading to the formation of an ``active foam'' like state. 
A different scenario for fingering has recently been studied in the context of chemotactic fronts \cite{Alert.etal2022}, where the front instability is mediated by a non-conserved chemoattractant field. Similar instabilities might be found in other recently studied systems where a phase-separating field is coupled to a non-conserved field through source terms \cite{Demarchi.etal2023,Zhao.etal2023}. 
Developing a capillary wave theory for the NRCH model is left for future work.
We expect that understanding the interface dynamics will be crucial to approach the open questions of coarsening and wavelength selection in 2D.

Third, we have restricted our analysis to the minimal model Eq.~\eqref{eq:NRCH} with constant coefficients. These minimal equations already give rise to unexpectedly rich phenomena, suggesting that many observations from experiments and more complex models might also be described by a minimal phenomenological model.
The simplicity of Eq.~\eqref{eq:NRCH} has allowed us to understand the dynamics in terms of linear stability analysis and phase space geometry, providing a starting point for systematic investigations of more complex models with density-dependent transport coefficients \cite{Saha.Golestanian2022} or spontaneous phase separation of the $\psi$-field \cite{Frohoff-Hulsmann.etal2021}.

A promising experimental setting for observing signatures of NRCH dynamics may be the \textit{in vitro} reconstituted Min-protein system of \textit{E.\ coli} \cite{Loose.etal2008}. Mathematical models of the Min system can be mapped to the NRCH equations since the densities of MinD and MinE are conserved and their interactions are nonreciprocal. The protein interactions can be tuned at a molecular level~\cite{Glock.etal2019a}.
Moreover, the Min-protein system can be confined to shallow microchambers \cite{Ivanov.Mizuuchi2010,Brauns.etal2021b}. This suppresses bulk-surface oscillations that appear in systems with larger bulk-surface volume \cite{Halatek.Frey2018}, allowing one to integrate out the bulk degrees of freedom.
Notably, in such microchambers, the Min system exhibits spatiotemporal chaos and traveling droplets not unlike those found in the NRCH model (see Movies~5 and~6). The insights from the NRCH model might also shed light on the quasi-stationary patterns of traveling waves observed in the Min-protein system and the transitions between such patterns~\cite{Glock.etal2019}.

Finally, the interface undulations also bear striking similarities to those found in phase separating mixtures of passive fluids and active liquid crystals driven out of equilibrium by cytoskeletal motor proteins~\cite{Adkins.etal2022}. In these systems it has been shown that the emergence of active emulsions of continuously splitting and merging droplets and of traveling interfacial waves is facilitated by the coupling to liquid crystalline degrees of freedom~\cite{Adkins.etal2022,Caballero.Marchetti2022}. The present work suggests, however, that variations of the purely scalar minimal model described here may also capture this behavior.

\acknowledgements{We thank Markus Bär, Jonathan Bauermann, Cesare Nardini and Zhihong You for insightful discussions. FB was supported by Simons Foundation grant (\#216179) to the KITP. MCM was supported by the National Science Foundation award No.\ DMR-2041459.}

\appendix

\section{Two-compartment approximation: mapping to FHN}
\label{app:mapping-to-FHN}

As a minimal ``cartoon'' of the spatially extended system, we can approximate the spatial extended domain by two diffusively-coupled well-mixed compartments \cite{Brauns.etal2021a}
\begin{subequations}
\begin{align*}
    \partial_t \phi_\mathrm{L} &= D_{11} (\phi_\mathrm{R} - \phi_\mathrm{L}) + \phi_\mathrm{L}^3 - \phi_\mathrm{R}^3 + D_{12} (\psi_\mathrm{R} - \psi_\mathrm{L}), \\
    \partial_t \phi_\mathrm{R} &= -\partial_t \phi_\mathrm{L}, \\
    \partial_t \psi_\mathrm{L} &= D_{21} (\phi_\mathrm{R} - \phi_\mathrm{L}) + D_{22} (\psi_\mathrm{R} - \psi_\mathrm{L}), \\
    \partial_t \psi_\mathrm{R} &= -\partial_t \psi_\mathrm{L}
\end{align*}
\end{subequations}
The average densities $\phi_0 = (\phi_\mathrm{L} + \phi_\mathrm{R})/2$ and $\psi_0 = (\psi_\mathrm{L} + \psi_\mathrm{R})/2$ are conserved. Therefore, the dynamics can be written in terms of the \emph{redistributed} densities $\Delta \phi = (\phi_\mathrm{R} - \phi_\mathrm{L})/2$, $\Delta \psi = (\psi_\mathrm{L} - \psi_\mathrm{R})/2$
\begin{subequations} \label{eq:FHN-mapped}
\begin{align}
    \tfrac{1}{2} \partial_{t} \Delta \phi &= -D_{11} (1 - 3 \phi_0^2) \Delta \phi - \Delta \phi^3 + D_{12} \Delta \psi, \\
    \tfrac{1}{2} \partial_{t} \Delta \psi &= D_{21} \Delta \phi - D_{22} \Delta \psi.
\end{align}
\end{subequations}
These equations have exactly the form of the FHN model Eq.~\eqref{eq:FHN} with ``offset'' $a = 0$.
This zero offset is a consequence of the parity symmetry of Eqs.~\eqref{eq:NRCH}. For spatially varying coefficients in Eq.~\eqref{eq:NRCH}, this parity symmetry is broken and one obtains an FHN equation with $a \neq 0$ in the two-compartment approximation.

Note that the two-compartment approximation corresponds to a system with no-flux boundary conditions, which exhibits standing waves rather than traveling waves. The continuous translational symmetry required for traveling waves can be captured by a single-mode approximation \cite{You.etal2020}.

\section{Unequal mixtures ($\bar{\phi} \neq 0$)}
\label{app:unequal-mix}

\begin{figure}
    \centering
    \includegraphics{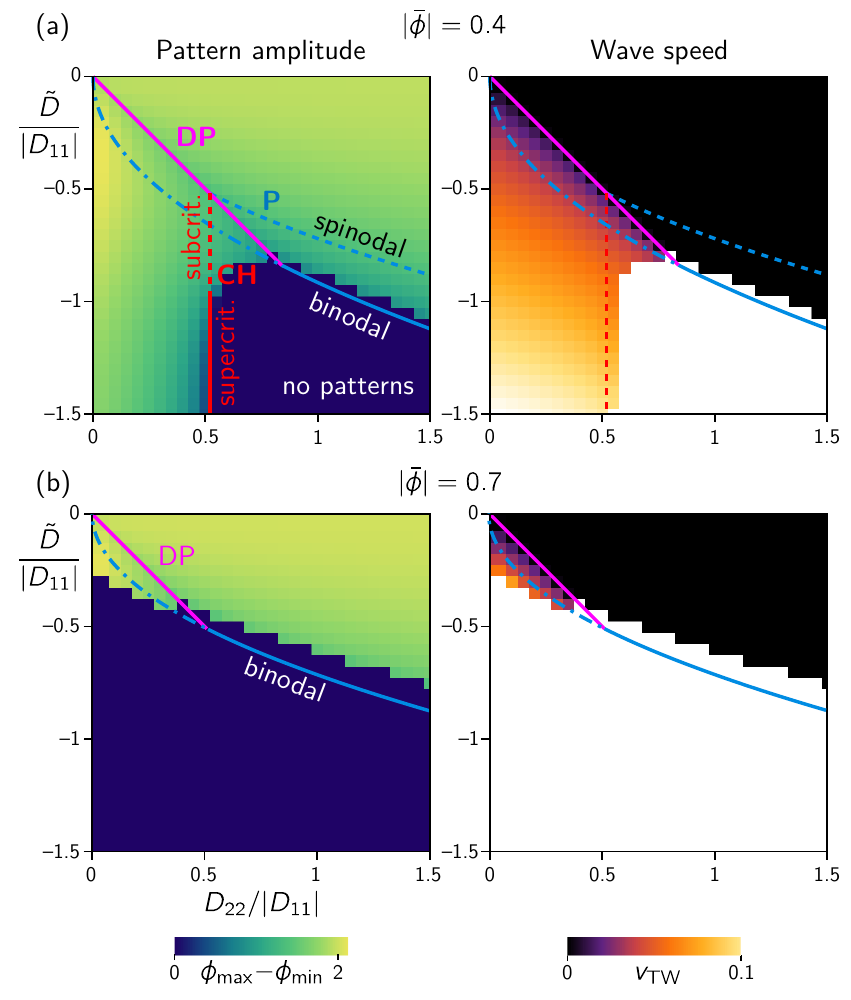}
    \caption{
    Phase diagrams for unequal mixtures ($\bar{\phi} \neq 0$).
    (a)~The homogeneous steady state is unstable above the spinodal (dashed blue line), while stationary patterns exist above the binodal (solid and dot-dashed blue line, the discrepancy to the numerical simulations is due to the finite size of the simulation domain).
    Below the drift-pitchfork bifurcation (solid magenta line, DP), stationary patterns are unstable and develop into traveling waves. 
    The red line indicates the conserved Hopf bifurcation (CH). Near the drift-pitchfork bifurcation, the conserved Hopf bifurcation is subcritical (dashed red line), while it is supercritical further away. 
    (b)~When $|\bar{\phi}|^2 > -D_{11}/3$, the homogeneous steady state is stable for all values of $D_{22}$ and $\tilde{D} < 0$, such that no spinodal and no conserved Hopf bifurcation appear in the phase diagram. Traveling waves emerging due to a drift-pitchfork bifurcation from stationary patterns exist for sufficiently weak anti-reciprocal coupling.
    }
    \label{fig:unequal-mix}
\end{figure}

In the main text, we have focused our analysis on the case of an equal (symmetric) mixture $\bar{\phi} = 0$. In this case, the onset of pattern formation is supercritical as the spinodal and the binodal lines coincide. In the following, we briefly discuss unequal mixtures.
As shown in Fig.~\ref{fig:unequal-mix}, the onset of pattern formation is generically subcritical when $\bar{\phi} \neq 0$.

In the phase diagram, the binodal lines $\bar{\phi} = \phi_\pm$, bounding the regime where stationary patterns exist, are given by $\tilde{D}_b = - \sqrt{(- D_{11} - \bar{\phi}^2) D_{22}}$.
The spinodal, bounding the regime where the homogeneous steady state is unstable, is given by $\tilde{D}_s = - \sqrt{(-D_{11} - 3\bar{\phi}^2) D_{22}}$. Because it patterns exist outside the spinodal region, the spinodal line is a \emph{subcritical} pitchfork bifurcation.
Notably, the locus of the exceptional point in the local dispersion relation [cf.\ Fig.~\ref{fig:LSA}] predicts precisely the drift-pitchfork bifurcation (DP) where traveling waves emerge from phase separated patterns.
This is true even deeply in the subcritical regime where the global homogeneous steady state is stable for all values of $\tilde{D}$ and $D_{22}$ [Fig.~\ref{fig:unequal-mix}(b)]. 

Near the drift-pitchfork bifurcation, the conserved Hopf bifurcation is subcritical while it remains supercritical further away [Fig.~\ref{fig:unequal-mix}(a)].

\section{Linear stability diagrams in the $D_-$-$D_+$ plane}
\label{app:Dplus-Dminus}

\begin{figure*}
    \centering
    \includegraphics{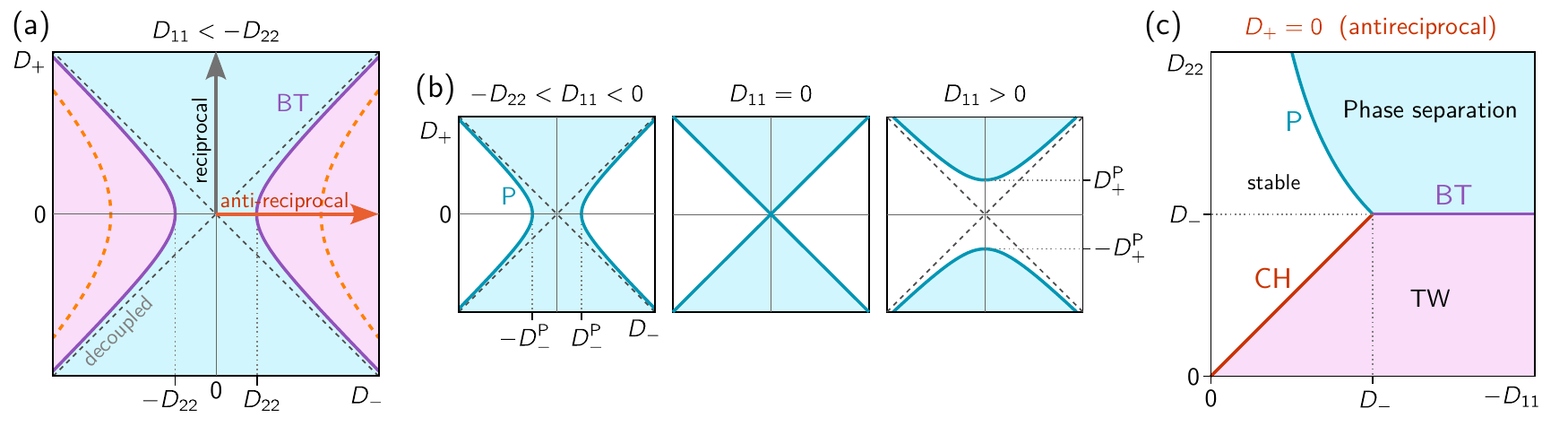}
    \caption{
    (a) and (b) Phase diagrams in the $D_-$-$D_+$ plane. The coordinate axes correspond to the reciprocal and anti-reciprocal cases, respectively. The bifurcation lines have hyperbolic shapes because only the product $D_{12} D_{21} = D_+^2 - D_-^2$ is relevant.
    (a) For sufficiently slow diffusion of $\psi$ (i.e.\ $D_{11} < -D_{22} < 0$), there is a regime of traveling waves.
    (b) No regime traveling waves exist when $D_{22} > -D_{11}$.
    For $D_{11} < 0$, sufficiently strong anti-reciprocal cross-diffusive ($D_- > D_-^\mathrm{p} = \sqrt{-D_{11} D_{22}}$) suppresses phase separation.
    For $D_{11} > 0$, the uncoupled system is linearly stable but sufficiently strong reciprocal cross-diffusive coupling ($D_+ > D_+^\mathrm{p} = \sqrt{D_{11} D_{22}}$) can destabilize the homogeneous steady state, giving rise to cross-diffusion induced static phase separation. 
    (c) Phase diagram in the $D_{11}$-$D_{22}$ plane in the anti-reciprocal regime ($D_+ = 0$). Note that this phase diagram shows the same information as the one in Fig.~\ref{fig:LSA}(a).
    }
    \label{fig:Dm-Dp-diagrams}
\end{figure*}
In this Appendix, we redraw our phase diagram in the plane of $D_\pm=D_{12}\pm D_{21}$ for the original form of Eq.~\eqref{eq:NRCH} where $\psi$ has not been rescaled. This parametrization has been used before in the literature \cite{Fruchart.etal2021,Saha.etal2020} and is therefore useful for making contact with previous work. In this parametrization the axes $D_-=0$ and $D_+=0$ correspond to purely reciprocal and purely anti-reciprocal cases, respectively [see Fig.~\ref{fig:Dm-Dp-diagrams}(a,b)]. Due to the freedom to rescale $\psi$, only the product $D_{12} D_{21} = D_+^2 - D_-^2$ controls the behavior of the system and therefore the bifurcation lines in the $D_\pm$ plane have a hyperbolic shape.
By rescaling $\psi$, one can always map system onto a purely reciprocal or purely anti-reciprocal one, depending on the sign of $D_{12} D_{21}$ as indicated by the black and red arrows, respectively.
In the regions between the solid purple and orange dashed lines systems with no-flux boundary conditions exhibit arrested traveling waves [cf.\ Fig.~\ref{fig:arrested-waves}(b)].
Note that large values of $D_{22}$, corresponding to fast diffusion of the $\psi$ field, allow the $\psi$ field to catch up with $\phi$, stabilizing the static phase separated state and suppressing the traveling waves (regions shaded in white in the phase diagrams).

Traveling wave patterns exist only for $D_{22} < -D_{11}$ [see Fig.~\ref{fig:Dm-Dp-diagrams}(a), cf.\ Fig.~\ref{fig:LSA}(a)]. Keeping $D_{22}$ fixed and decreasing the negative value of $D_{11}$ thus eliminates traveling waves from the phase diagram once $-D_{11} < D_{22}$.

\section{Arrested traveling waves and standing waves}
\label{app:no-flux}

\begin{figure*}
    \centering
    \includegraphics{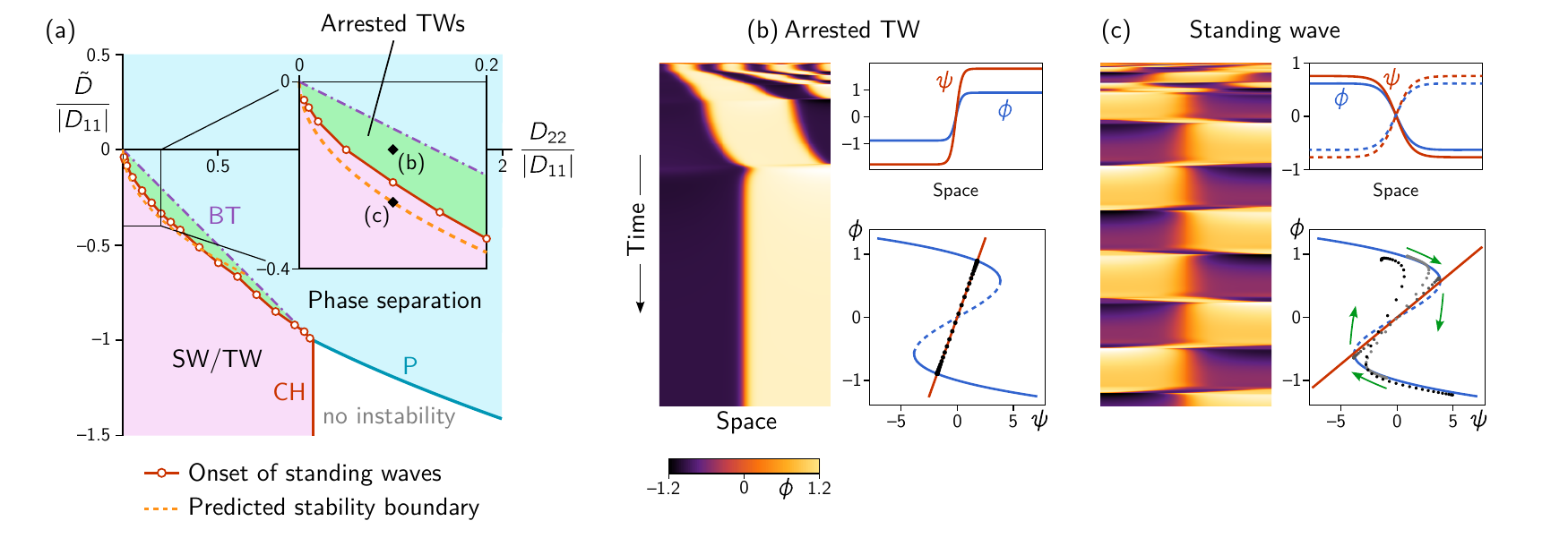}
    \caption{
    (a)~Phase diagram for a 1D domain with no-flux boundary conditions arrests the drift-pitchfork bifurcation. The regime of arrested traveling waves found in numerical simulations is shaded in green. For sufficiently strong anit-reciprocal coupling (below the red line), arrested traveling waves transition to standing waves due to a destabilization of the bulk domains. Along the dashed orange line, the $\psi$-nullcline intersects the $\phi$-nullcline at it's turning points [see (c)]. In the timescale-separated regime $D_{22} \ll |D_{11}|$, this marks the limit of existence of stable stationary patterns.
    (b)~Arrest of the interface-driven traveling waves once all mass has accumulated at a no-flux boundary ($D_{12} = -D_{21} = 0.14$). Note that the nullcline intersection points corresponding to the bulk densities lie on the stable segments (solid blue line) of the $\phi$-nullcline in the phase portrait.
    (c)~Standing wave pattern resulting from the destabilization of the bulk domains ($D_{12} = -D_{21} = 0.25$).
    (Parameters: $D_{11} = -1, D_{22} = 0.1, L = 50$, simulation time $t = 0 - 2{\times}10^4$.)
    }
    \label{fig:arrested-waves}
\end{figure*}

In the main text, we have argued that traveling waves are driven by the propagating interface mode and that they require continuous mass transport through the bulk domains. 
No-flux boundary conditions suppress this mass transport and thereby stabilize a stationary interface against the drift-pitchfork bifurcation; see Fig.~\ref{fig:arrested-waves}.
In 2D, such ``arrested'' interfaces are destabilized by mass transport transversal to the interface [cf.\ Fig.~\ref{fig:2D-patterns}(d)].
In 1D, the stationary patterns become unstable only when the bulk domains lose stability.
An approximate criterion for this instability can be read-off from the phase-portrait construction, which suggests that the bulk domains are unstable when $(\phi_\pm, \psi_\pm)$ lie on the positively sloped segment of the $\phi$-nullcline.
In other words, stationary patterns are expected to lose stability when the $\psi$-nullcline intersects the $\phi$-nullcline at its turning points.
Indeed, this criterion provides a good approximation for the onset of standing wave oscillations; see Fig.~\ref{fig:arrested-waves}(a,c).
It is analogous to the geometric criterion for the onset of limit cycle oscillations in the FHN model [cf.\ Fig.~\ref{fig:FHN}(b)].

\section{Numerical simulations}
\label{app:numerics}

1D simulations were performed using \emph{Mathematica's} \texttt{NDSolve[]} function. 2D Simulations were performed using the finite element software \emph{COMSOL Multiphysics v6}. 
Since finite element methods are not well-suited to handle gradients beyond second order, we introduce $\mu_\phi$ as an auxiliary field and write the $\phi$ equation as $\partial_t \phi = \nabla^2 \mu_\phi$. To make the equation for $\mu_\phi$ parabolic, we allow it to relax on a fast timescale $\tau_\mu \ll 1$: 
\begin{equation}
    \tau_\mu \partial_t \mu_\phi + \mu_\phi = D_{11} \phi + \phi^3 + \tilde{D} \psi - \kappa \nabla^2 \phi.
\end{equation}
We used $\tau_\mu = 10^{-4}$. Choosing smaller values of $\tau_\mu$ did not change the results.

The wave speed in 1D simulations was calculated via
\begin{equation}
    v_\mathrm{TW} = -\frac{\int_0^L \mathrm{d} x \, \partial_x \phi \, \partial_t \phi}{\int_0^L \mathrm{d} x \, (\partial_x \phi)^2},
\end{equation}
which robustly determines the average speed of moving interfaces where $(\partial_x \phi)^2$ is large as one can easily see by transforming into the co-moving frame $\phi(x,t) = \tilde{\phi}(x - v_\mathrm{TW} t)$.

\begin{figure}[t]
    \centering
    \includegraphics{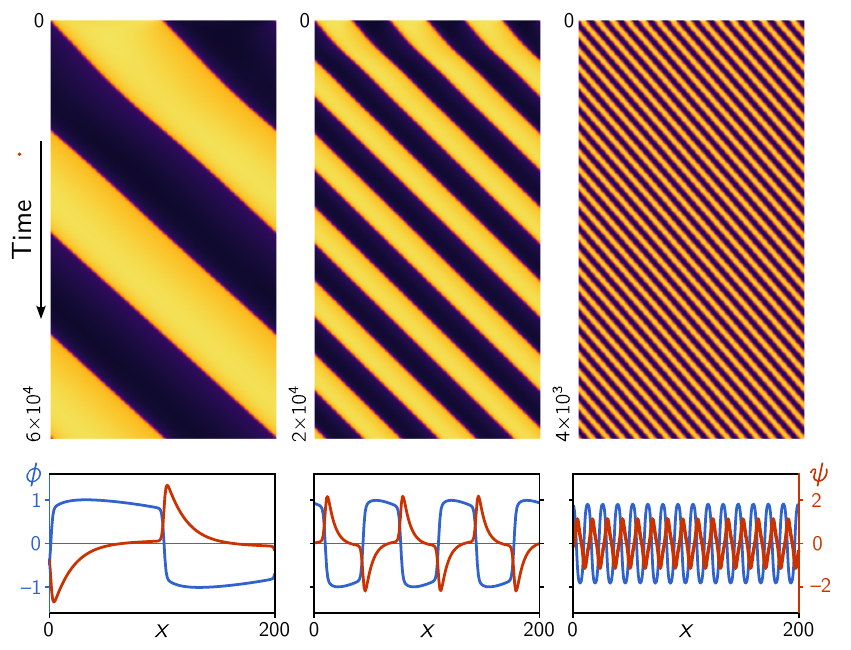}
    \caption{Multistability of traveling waves initiated with different wavelengths $\lambda = L/n$ with $n = 1, 3, 15$ the number of ``droplets''. Note that the time axis is scaled differently in each kymograph to account for the scaling of wavespeed with inverse wavelength, $v_\mathrm{TW} \sim 1/\lambda$. Parameters: $L = 200, D_{22} = 0.1, D_{12} = 0.2$.}
    \label{fig:TW-multistability}
\end{figure}

To asses patterns in the binodal regime [e.g.\ in Fig.~\ref{fig:wave-speed}(d) and Fig.~\ref{fig:unequal-mix}], simulations were initialized in a phase-separated state
\begin{align}
    \phi(x,0) &= \tanh\!\left[ \left(\tfrac{L_+}{2} - |x - \tfrac{L}{2}|\right) / \ell \right] + \xi_1(x), \\
    \psi(x,0) &= \sin(2 \pi x/L) + \xi_2(x).
\end{align}
Here, $L$ is the simulation domain size, $L_+ = (1 + \bar{\phi}/2)$ is the size of the high-density domain, and $\ell$ controls the initial width of interfaces.
$\xi_{1,2}$ denote small amplitude noise (uniformly distributed in $[-0.1, 0.1]$).
This initial condition, extended to 2D was also used to study the emergence of interface undulations in 2D [cf.\ Fig.~\ref{fig:2D-patterns}(b--d)].

To assess the stability of traveling waves with different wavelengths, we performed simulations starting with initial conditions given by 
\begin{align}
    \phi(x,0) &= \tanh\bigl[\sin(2 \pi n x/L)/\ell\bigr] + \xi_1(x), \\
    \psi(x,0) &= \cos(2 \pi n x/L) + \xi_2(x).
\end{align}
The $\pi/2$ phase shift between $\phi$ and $\psi$ sets the direction of wave propagation.
We found that all wavelengths above the interface width are stable, including the fully phase separated state ($n = 1$).
Kymographs of three examples are shown in Fig.~\ref{fig:TW-multistability}.

\begin{figure}
    \centering
    \includegraphics{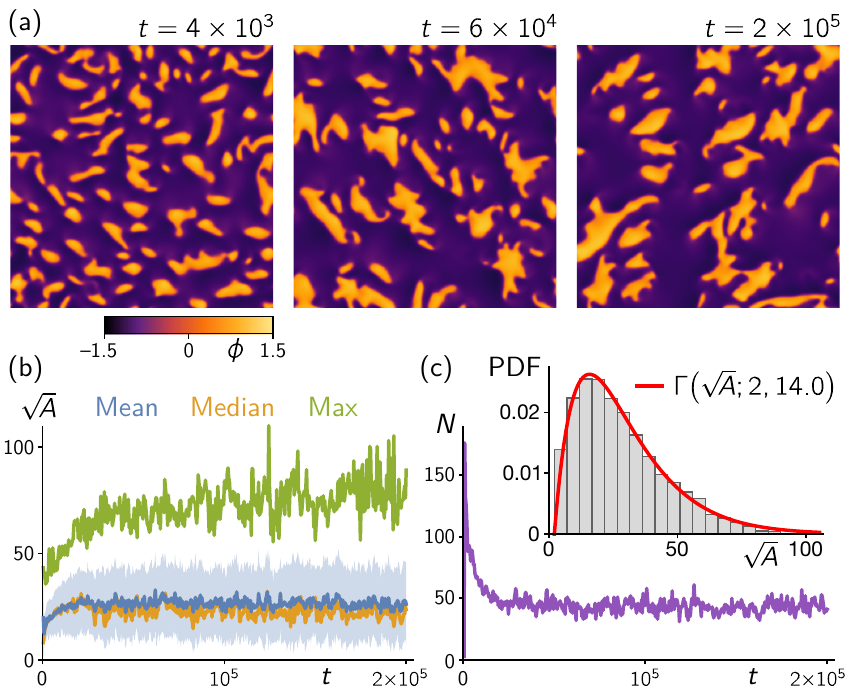}
    \caption{
    (a)~Snapshots of a long-term simulation showing interrupted coarsening of traveling droplets.
    The droplet areas (b) and number (c) continue to fluctuate as droplets continually collide and break up (shaded band in indicates one standard deviation around the mean). 
    The distribution of droplet sizes [see inset in (c)] reaches a steady state that appears to be well approximated by a Gamma distribution with shape parameter $\alpha = 2$ and scale parameter $\beta = 14$, i.e.\ $P(\sqrt{\!A}) \sim \sqrt{\!A} \exp[\sqrt{\!A}/14.0]$. Droplet sizes were pooled across the time interval $t \in [0.6,2]{\times} 10^5$.
    }
    \label{fig:droplet-size-distribution}
\end{figure}

We investigated the interrupted coarsening dynamics of droplets in a system with non-equal mixture ($\bar{\phi} = -0.5$) by running a simulation until the distribution of droplet sizes reached a steady state (see Fig.~\ref{fig:droplet-size-distribution}).

\section{Mass-conserving reaction--diffusion systems} \label{app:MCRD}

\subsection{General mass-conserving reaction--diffusion systems}

Dynamical equations describing reaction--diffusion can be written in the general form
\begin{equation} \label{eq:RD-general}
    \partial_t \uvec(x,t) = \mathbf{D} \nabla^2 \uvec + \fvec(\uvec)\,,
\end{equation}
where $\uvec=\{u_i\}$ represents the collection of all diffusing and reacting densities and $\mathbf{D} = \diag(\{D_i\})$ is the  diffusion matrix. We demand the diffusion matrix be diagonal, meaning that there is no cross-diffusion on the level of the individual components $u_i$.
When the reaction kinetics $\fvec$ is mass conserving, there exist ``stochiometric'' vectors $\mathbf{s}_i$ such that $\mathbf{s}_i^\mathrm{T}\fvec = 0$. The corresponding conserved densities are given by $\rho_i = \mathbf{s}_i^\mathrm{T}\uvec$.
We can further define the mass-redistribution potentials $\eta_i(\uvec) := \mathbf{s}_i^\mathrm{T}\mathbf{D}\uvec$.
Multiplying Eq.~\eqref{eq:RD-general} with $\mathbf{s}_i^\mathrm{T}$ from the left then yields the mass-redistribution dynamics
\begin{equation}
    \partial_t \rho_i(x,t) = \nabla^2 \eta_i(\uvec(x, t))\,.
\end{equation}

To find approximate, closed dynamics for the densities $\rho_i$, we introduce the local reactive equilibria \cite{Halatek.Frey2018,Brauns.etal2020}, which are functions of the local total densities $\rho_i$
\begin{equation} \label{eq:loc-eq}
    \uvec^*(\{\rho_i\}) \; : \;
    \left\{ \begin{array}{rl}
        \fvec(\uvec^*) =& \!\! 0, \\
        \mathbf{s}_i^\mathrm{T}\uvec^* =& \!\! \rho_i \quad \forall i.
    \end{array} \right.
\end{equation}
Under the condition that the local reactive dynamics $\partial_t \uvec = \fvec(\uvec)$ have a single stable fixed point $\uvec^*$, we can make a local equilibrium approximation in the long wavelength limit
\begin{align}
	\uvec(x, t) \approx \uvec^*\boldsymbol{(}\{\rho_j(x,t)\}\boldsymbol{)} \, ,
\end{align}
and reduce the dynamics to the redistribution of the total densities
\begin{equation}
    \partial_t \rho_i(x,t) = \partial_x^2 \eta_i^*(\{\rho_j\}) =
    \partial_x \left[\sum_{j} \mathcal{D}_{ij} \, \partial_x \rho_j  \right]\,,
\end{equation}
with the effective (cross-)diffusion coefficients $\mathcal{D}_{ij} := \mathbf{s}^\mathrm{T}_i \mathbf{D} \partial_{\rho_j} \uvec^*$.
Effective cross-diffusion of the conserved densities arises as a consequence of the chemical reactions even when the ``bare'' diffusion matrix $\mathbf{D}$ is diagonal.
The coefficients $\mathcal{D}_{ij}$ will in general depend on the local densities, $\rho_i$, via the local equilibrium concentrations $\uvec^*(\{\rho_i\})$.

When the determinant of the effective diffusion matrix $\mathcal{D}$ is negative, the dynamics becomes effectively anti-diffusive, i.e.\ has a long-wavelength instability. At short wavelengths, the local equilibrium approximation doesn't hold and the system is restabilized by the interplay between (slow) diffusion and reactions \cite{Brauns.etal2020}. Accounting for this interplay in the local equilibrium approximation yields the stabilizing $\nabla^4$ term in \eqref{eq:NRCH-phi}.

\subsection{Min-protein dynamics}

The ``skeleton model'' of the Min system \cite{Halatek.Frey2012,Brauns.etal2021b} describes the concentrations of MinD and MinE in different conformational states, namely membrane-bound MinD ($\md$) and MinDE complexes ($\mde$), and cytosolic MinD-ATP, MinD-ATP, and MinE ($\cDT, \cDD, \cE$).
The model can be written in the form Eq.~\eqref{eq:RD-general} with
$\uvec = (\md, \mde, \cDT, \cDD, \cE)$, $\mathbf{D} = \diag (\Dd, \Dde, \DD, \DD, \DE)$ and the reactive dynamics are given by
\begin{equation} \label{eq:Min-reactions}
	\fvec(\uvec) = 
	\begingroup \renewcommand*{\arraystretch}{1.2} \begin{pmatrix}
	\phantom{-}R_\mathrm{D}^\mathrm{on} (\uvec) - R_\mathrm{E}^\mathrm{on} (\uvec) \\
	\phantom{-}R_\mathrm{E}^\mathrm{on} (\uvec) - R_\mathrm{DE}^\mathrm{off} (\uvec) \\
	-R_\mathrm{D}^\mathrm{on} (\uvec) + \lambda \cDD \\
	R_\mathrm{DE}^\mathrm{off} (\uvec) - \lambda \cDD \\
	-R_\mathrm{E}^\mathrm{on} (\uvec) + R_\mathrm{DE}^\mathrm{off} (\uvec)
	\end{pmatrix} \endgroup,
\end{equation}
where the reaction terms
\begin{subequations}
	\begin{align}
		R_\mathrm{D}^\mathrm{on} (\uvec) &= (\kD + \kdD\md)\cDT \, , \\
		R_\mathrm{E}^\mathrm{on}(\uvec) &= \kdE\md\cE \, , \\
		R_\mathrm{DE}^\mathrm{off}(\uvec) &= \kde\mde \, ,
	\end{align}
\end{subequations}
account, respectively, for MinD attachment and self-recruitment to the membrane, MinE recruitment by MinD, and dissociation of MinDE complexes with subsequent detachment of both proteins to the cytosol. 
The term $\lambda \cDD$ accounts for nucleotide exchange, i.e.\ conversion from $\cDD$ to $\cDT$, in the cytosol.
Importantly, these reaction kinetics conserve the average total density of MinD and MinE proteins, $\bar{\rho}_\mathrm{D}^{}$ and $\bar{\rho}_\mathrm{E}^{}$, individually, i.e.\ there are two globally conserved masses that are redistributed in space.

\section{Active gel models} \label{app:poroelastic}

\subsection{Mapping to effective NRCH equation}
To derive Eq.~\eqref{eq:active-gel-effective} from the 1D active gel model Eqs.~\eqref{eq:active-gel}, we first rewrite the force balance equation as
\begin{equation}
    (\gamma - \eta \partial_x^2) \dot{u} = E \, \partial_x^2 u + \partial_x T_\mathrm{a}(c)
\end{equation}
and then formally solve for $\dot{u}$. The resulting kernel $(\gamma - \eta \partial_x^2)^{-1}$ can be approximated in a gradient expansion, yielding (to third order in $\partial_x$)
\begin{align}
    \gamma \dot{u} &\approx \left(1 + \tfrac{\eta}{\gamma} \partial_x^2 \right) \left[E \, \partial_x^2 u + \partial_x T_\mathrm{a}(c) \right] \\
    &\approx E \, \partial_x^2 u + \partial_x T_\mathrm{a}(c) + \frac{\eta}{\gamma} \partial_x^3 T_\mathrm{a}(c).
\end{align}
After applying $\partial_x$ on both sides and introducing the strain $\varepsilon = \partial_x u$ one has 
\begin{equation}
    \gamma \partial_t \varepsilon \approx E \, \partial_x^2 \varepsilon + \partial_x^2 T_\mathrm{a}(c) + \frac{\eta}{\gamma} \partial_x^4 T_\mathrm{a}(c).
\end{equation}
Substituting $\dot{u}$ into Eq.~\eqref{eq:cell-density} gives
\begin{align}
    \gamma \partial_t c = \gamma \partial_x^2 D - \partial_x \! \left[ c E \, \partial_x^2 u + \partial_x T_\mathrm{a}(c) \right] - \frac{\eta}{\gamma} \partial_x^4 T_\mathrm{a}(c).
\end{align}
In Eq.~\eqref{eq:active-gel}, we have retained the fourth order gradient only in the $c$-dynamics, where it is necessary to stabilize short lengthscales when $T_a' > \gamma D/c$. 

\subsection{Active poroelastic models}

As we briefly explain in the following, the models by Radszuweit \textit{et al.}\ \cite{Radszuweit.etal2013} and by Weber \textit{et al.}\ \cite{Weber.etal2018} are closely related. This fact is slightly obscured by the non-dimensionalization performed in Weber \textit{et al.}\ which uses the elasticity constant to rescale time and thus prevents decoupling the displacement field from the mass-redistribution dynamics in the limit of a vanishing elastic constant. The explicit concentration field $c$ in \cite{Radszuweit.etal2013} is replaced by the volume fraction in \cite{Weber.etal2018}. Both models can be cast in the form of Eq.~\eqref{eq:active-gel}.

We recapitulate the derivation from \cite{Weber.etal2018}, immediately specifying to an active solid with volume fraction $\phi$ and a passive fluid with volume fraction $1 - \phi$. The dynamics of $\phi$ is governed by a transport equations and an incompressibility equation 
\begin{align}
    \partial_t \phi + \partial_x (v_\mathrm{s} \phi) = D \partial_x^2 \phi\,, \label{eq:solid-transport} \\
    0 = \partial_x( \phi v_\mathrm{f} + (1 - \phi) v_\mathrm{s})\,.
\end{align}
Given appropriate boundary conditions (Dirichlet or periodic) the incompressibility condition can be solved to give $v_\mathrm{f} = -\phi/(1-\phi) v_\mathrm{s}$.

Force balance in the two phases reads
\begin{align}
    0 &= \partial_x(\phi \sigma_\mathrm{s}) - \phi \partial_x p - f\,, \label{eq:force-balance-solid} \\
    0 &= \partial_x[(1-\phi) \sigma_\mathrm{f}] - (1-\phi) \partial_x p + f\,, \label{eq:force-balance-fluid}
\end{align}
with the friction force $f = \gamma \phi (1 - \phi) (v_\mathrm{s} - v_\mathrm{f})$. The pressure $p$ serves as a Lagrange multiplier to enforce incompressibility. The active solid is described by a Kelvin--Voigt model together with an active tension
\begin{equation}
    \sigma_\mathrm{s} = E \partial_x u + \eta \partial_x v_\mathrm{s} + T_\mathrm{a}(\phi)\,,
\end{equation}
where $u$ is the solid's displacement, i.e.\ $\dot{u} = v_\mathrm{s}$.
The viscous stress in the fluid phase is negligible compared to the solid stresses, so we can set $\sigma_\mathrm{f} = 0$. Solving Eq.~\eqref{eq:force-balance-fluid} for $\partial_x p$ and substituting into Eq.~\eqref{eq:force-balance-solid} then gives
\begin{equation}
    \frac{\gamma \phi^2}{1-\phi} \dot{u} - \eta \, \partial_x (\phi \partial_x \dot{u}) = E \partial_x (\phi \partial_x u) + \partial_x (\phi T_\mathrm{a}(\phi))\,. \label{eq:force-balance-Weber}
\end{equation}
Together with Eq.~\eqref{eq:solid-transport}, we have a closed system of equations for the dynamics of $\phi$ and $u$.

Note the close similarity to Eq.~\eqref{eq:force-balance}. In fact, for small deviations from a uniform state $\phi = \phi_0$, we can approximate $\phi$ as constant in all terms in \eqref{eq:force-balance-Weber} except for the last one which is responsible for the instability. 

The origin of the friction term in the viscoelastic gel case Eq.~\eqref{eq:force-balance} and the poroelastic case Eq.~\eqref{eq:force-balance-Weber} differ. In the former case, it accounts for friction with a rigid substrate, whereas in the latter case, it accounts for friction between the two interpermeating phases.

%

\end{document}